\documentclass[a4paper,fleqn,usenatbib]{mnras}

\usepackage[T1]{fontenc}
\usepackage{ae,aecompl}

\usepackage{graphicx}	
\usepackage{amsmath}	
\usepackage{amssymb}	
\usepackage{ulem}		

\title[C, $^{12}$C/$^{13}$C and N in solar twins]{Carbon, isotopic ratio $^{12}$C/$^{13}$C and nitrogen in solar twins:
constraints for the chemical evolution of the local disc}

\author[R. B. Botelho et al.]{R. B. Botelho,$^{1}$\thanks{E-mail: rafaelbuenobotelho@gmail.com (RBB)}
A. de C. Milone,$^{1}$\thanks{E-mail: andre.milone@inpe.br (ADCM)}
J. Mel{\'e}ndez,$^{2}$
A. Alves-Brito,$^{3}$
L. Spina$^{4,5}$
\newauthor and J. L. Bean$^{6}$
\\$^{1}$Divis\~ao de Astrof\'\i sica, Instituto Nacional de Pesquisas Espaciais, Av. dos Astronautas 1758, S\~ao Jos\'e dos Campos, 12227-010, Brazil
\\$^{2}$Departamento de Astronomia, IAG, Universidade de S\~ao Paulo, Rua do Mat\~ao 1226, S\~ao Paulo, 05509-900, Brazil
\\$^{3}$Instituto de F\'\i sica, Universidade Federal do Rio Grande do Sul, Av. Bento Gon\c calves 9500, Porto Alegre, RS, Brazil
\\$^{4}$Monash Center for Astrophysics, School of Physics and Astronomy, Monash University, Clayton, VIC 3800, Australia
\\$^{5}$Australian Research Council, Centre of Excellence in All Sky Astrophysics in 3 Dimensions, Canberra, Australia
\\$^{6}$Department of Astronomy \& Astrophysics, University of Chicago, 5640 S. Ellis Ave, Chicago, IL 60637, USA
\\
}

\date{Accepted XXX. Received YYY; in original form ZZZ}

\pubyear{2020}

\begin{document}
\label{firstpage}
\pagerange{\pageref{firstpage}--\pageref{lastpage}}
\maketitle

\begin{abstract}
Abundances of light elements in dwarf stars of different ages are important constraints
for stellar yields, Galactic chemical evolution and exoplanet chemical composition studies.
We have measured C and N abundances and $^{12}$C/$^{13}$C ratios
for a sample of 63 solar twins spanning a wide range in age,
based on spectral synthesis of a comprehensive list of CH\,A-X and CN\,B-X features
using HARPS spectra.
The analysis of 55 thin disc solar twins confirms the dependences of [C/Fe] and [N/Fe] on [Fe/H].
[N/Fe] is investigated as a function of [Fe/H] and age for the first time for these stars.
Our derived correlation [C/Fe]-age agrees with works for solar-type stars and solar twins,
but the [N/Fe]-age correlation does not.
The relations [C,N/Fe]-[Fe/H] and [C,N/Fe]-age for the solar twins lay under-solar.
$^{12}$C/$^{13}$C is found correlated with [Fe/H]
and seems to have decreased along the evolution of the local thin disc.
Predictions from chemical evolution models for the solar vicinity
corroborate the relations [C,N/Fe]-[Fe/H], $^{12}$C/$^{13}$C-age and [N/O]-[O/H], 
but do not for the $^{12}$C/$^{13}$C-[Fe/H] and [C/O]-[O/H] relations.
The N/O ratio in the Sun is placed at the high end of the homogeneous distribution of solar twins,
which suggests uniformity in the N-O budget
for the formation of icy planetesimals, watery super-earths and giant planets.
C and N had different nucleosynthetic origins along the thin disc evolution,
as shown by the relations of [C/N], [C/O] and [N/O] against [O/H] and age.
[C/N] and [C/O] are particularly observed
increasing in time for solar twins younger than the Sun.
\end{abstract}




\section{Introduction}

A key aspect of the formation of stars and their planets regards to
the pristine chemical composition of their parental cloud of gas and dust.
Various connections between the chemical composition of planet-host stars and the planet frequency are observed and proposed
such as primordial abundance anomalies, planet engulfment events, retention of refractory elements by rocky planets and/or core of giant planets,
and due to the formation of Jupiter analogs,
which could trap significant quantity of dust exterior to their orbits preventing large amounts of refractory elements from being accreted to the forming star
\citep{santos2001, santos2004, fischer, Melendez2009, ramirez2011, Wang2015, Ghezzi2018, Petigura2018, Teske2019, Booth2020}.

The abundances of carbon, nitrogen and oxygen
beyond the ice line in a protoplanetary disc of gas/dust (or simply nebula)
are strictly related to the efficiency on the assembling of planetesimals
to form planetary embryos (bodies with sizes between the Moon and Mars) and cores of giant planets.
This is because at large distances from the central protostar,
the icy planetesimals existent there due to lower temperatures have their mutual sticking process enhanced
\citep{Haghighipour2013, Marboeufetal2014}.
By developing a model that computes the chemical composition and the ice-to-rock mass ratio in icy planetesimals
throughout protoplanetary discs of different masses, \citet{Marboeufetal2014} derived that
the ice content of these bodies is mainly dominated by H$_{2}$O, CO, CO$_{2}$, CH$_{3}$OH, and NH$_{3}$.
Their models predict the ice-to-rock ratio in icy planetesimals to be equal to 1($\pm$0.5), like observed in the Solar System's comets,
but still 2-3 times lower than usually considered in planet formation models.
Therefore, in some sense, the pristine content of CNO in a star plays an important role for the potential planetary formation around it,
and this could also be imprinted
in the chemical composition and structure of the formed planets (and vice-versa).

The chemical evolution of the solar neighbourhood is based on the stellar yields of all elements and other key ingredients,
such as the star formation rate, the matter infall from the halo,
the radial gas and star flows across the disc,
the vertical gas outflows from the disc,
the stellar initial mass function, the relative rates of different kinds of supernovae
and the frequency of close binaries
\citep{Tinsley1980, Pagel2009, Matteucci2012}.
The production of C, N, O and Fe along time, as well as of other abundant elements,
will determine the temporal evolution of their abundances in the interstellar medium (ISM),
and will be imprinted in the composition of stars and their planets.

The CNO surface abundances in evolved stars are modified
by the nuclear burning together with internal mixing processes
\citep{Charbonnel1994, Gratton2000, Maas2019}.
Therefore, dwarf stars are fundamentally indispensable to be observed in order to trace the CNO enrichment in the ISM.

Whilst carbon is a primary element (i.e., it is synthesized starting from H and He in the parent star),
nitrogen is considered a primary element (when produced, e.g., in low-metallicity fast-rotating massive stars, see \citet{Meynet2002a})
as well as a secondary one (when synthesized from $^{12}$C or $^{16}$O nuclei already present since the star birth).
The intermediate-mass (2-3\,$\leq$\,M\,$\leq$\,7-9\,M$_{\odot}$) and massive (M\,$\geq$\,7-9\,M$_{\odot}$) stars
are the C and N main sources to pollute the ISM
due to their shorter lifetimes relative to low-mass stars (M\,$\leq$\,2-3\,M$_{\odot}$),
although 1-2\,M$_{\odot}$ stars may also eject some amount of C and N
through the first thermal pulse on the asymptotic giant branch, AGB \citep{Karakas2016}.
The relative contributions between intermediate-mass and massive stars for the C and N productions are still uncertain
\citep{Meynet2002a, Meynet2002b}.

The carbon main isotope $^{12}$C is essentially made by the triple-$\alpha$ process.
The isotope $^{13}$C is residually made by different ways \citep{Meynet2002b}:
{\bf (a)} in the CN cycle also named as CNO-I cycle (H burning in intermediate-mass and massive stars),
{\bf (b)} in the $^{12}$C burning (low-metallicity fast-rotating massive stars),
and {\bf (c)} through proton-capture nucleosynthesis on the AGB
(intermediate-mass stars with hot bottom burning just below the convective envelope).

Nitrogen basically comes from C and/or O,
becoming its production more complex than carbon.
The most abundant isotope $^{14}$N is synthesized and ejected to the ISM through distinct ways:
{\bf (i)} in the complete CNO-II cycle
and ejected through the thermal pulses on the AGB phase for isolated stars only \citep{Pettini2002};
{\bf (ii)} through the $^{12}$C burning in low-metallicity fast-rotating massive stars
and ejected by their strong winds \citep{Meynet2002a};
{\bf (iii)} through a proton-capture reaction in the hot base of the convective envelope
and ejected during the AGB phase of intermediate-mass stars (like $^{13}$C) \citep{Marigo2001, Meynet2002a};
and {\bf (iv)} in the CN cycle from $^{12}$C and ON cycle from $^{16}$O in stars of any mass and ejected by them.
However, the relative importance of these processes is not very well determined.

Oxygen -- an $\alpha$-element -- is mainly ejected to the ISM in
the pre-supernova phase of the type II supernovae core collapse events (SN-II)
\citep{Chiappini2003}.
On the chemical enrichment scenario of the solar neighbourhood,
iron peak elements (Fe as a template) come
from a combinative production by type Ia supernovae (SN-Ia) and SN-II.
In fact, the observed decrease of [$\alpha$-element/Fe] versus [Fe/H]
in the thick and thin disc stars reflects the relative importance of SN-Ia over SN-II along the formation history of the Galaxy's disc
\citep{Feltzingetal2003, Bensbyetal2003, Melendezetal2008, Alves-Britoetal2010, RecioBlancoetal2014, Weinbergetal2019}.

Explosions of classical novae in close binary systems specifically play an important role on
the production of the rare CNO isotopes $^{13}$C, $^{15}$N and $^{17}$O \citep{Romano2019, JoseHernanz2007}.
On the other hand, SN-Ia events eject negligible amounts of CNO elements to the ISM.

Therefore, the ratios C/Fe, N/Fe, $^{12}$C/$^{13}$C, C/N, C/O and N/O in dwarf stars
provide valuable information about the main nucleosynthetic processes for these elements over the evolution of the Galaxy's disc.
For instance, the relative contributions of yields of single massive stars, yields of isolated AGB stars (low and intermediate-mass stars),
SN-II/SN-Ia ratio and close binary fraction could be understood
through robust Galactic chemical evolution (GCE) models
with those observations as constraints \citep{Chiappini2003, Kobayashi2011, Sahijpal2013, Romano2017, Romano2019}.
Specifically, \citet{Romano2017} states that measurements of $^{12}$C/$^{13}$C in wide samples of nearby dwarfs
would be very useful for construction of representative GCE models.

Analysing nearby solar twins provide an opportunity for studying
the chemical evolution of the local disc with time around the solar metallicity,
because they can encompass very different ages ranging since about the formation of the thin disc until now \citep{TucciMaia2016}.
The opportunity becomes more interesting when the solar twins have very well-determined fundamental parameters,
resulting in precise relative ages.
\citet{Bedell2018}, \citet{Spina2018} and \citet{Nissen2015}
have recently investigated how abundance ratios [X/Fe] in solar twins are related to age and [Fe/H].
\citet{Bedell2018} and \citet{Nissen2015}, for instance, similarly obtained that [C/Fe] and [O/Fe] increase with the stellar age,
but they did not include nitrogen and $^{12}$C/$^{13}$C in their studies.

We point out that nitrogen has not been derived for the current sample of solar twins yet,
albeit a first analysis of N in 11 solar twins was performed by \citet{Melendez2009}.
The main reason is due to a lack of measurable atomic lines in the optical region
and also because absorption features of molecules
(e.g. CH, NH and CN) are actually spectral blends requiring spectral synthesis.
For instance, equivalent widths of a weak atomic N\,I line at 7468.3\,{\AA} and spectral synthesis of NH in the ultraviolet
were used in the study of solar-type stars by \citet{Ecuvillon2004}.
More recently, \citet{DaSilva2015} and \citet{Suarez-Andres2016}, handling samples of solar-type dwarfs,
found that giant planet hosts are nitrogen-rich (in the [N/H] scale)
when compared with their control samples of dwarfs without known planets.
However, the increase of [N/Fe] as a function of [Fe/H]
and the negative trend of [N/Fe] with the stellar age, as observed in both works,
do not imply on any significant difference between the two stellar groups.

In the current work we have homogeneously measured precise abundances of carbon and nitrogen
plus the carbon main isotopic ratio $^{12}$C/$^{13}$C,
in a sample of well-studied solar twins, having distances up to 100\,pc and spanning a wide range of ages
since around the formation epoch of the Galaxy's thin disc until a few hundreds Myr ago \citep{Bedell2018, Spina2018, Botelho2019}.
The analysis of our results is focused on the distributions of a set of CNO abundance ratios
as a function of the age and metallicity of these solar twins 
in order to provide unprecedented and self-consistent constraints for GCE modelling \citep[e.g.][]{Romano2017, Sahijpal2013}.

Our paper is organized as follows. In Section 2, we present the sample of solar twins, their main parameters and the spectroscopic data.
Section 3 details the homogeneous chemical analysis carried out differentially to the Sun
for deriving the C abundance, the $^{12}$C/$^{13}$C isotopic ratio and the N abundance
based on the spectral synthesis of molecular features of the electronic systems CH\,A-X, CH\,A-X and CN\,B-X, respectively.
In Section 4, we study the behaviour of [C/Fe], [N/Fe], $^{12}$C/$^{13}$C, [C/N], [C/O] and [N/O]
as a function of [Fe/H] and isochrone stellar age,
and also [C/N], [C/O] and [N/O] as a function of [O/H] (other stellar metallicity indicator).
Finally, Section 5 summarizes our main results and conclusions.

\section{Solar twins sample and data}

The sample is composed of 67 solar twins, stars with 
atmospheric parameters very similar to the Sun, within around $\pm100\thinspace$K in 
$T_{\rm eff}$ and $\pm0.1\thinspace$dex in $\log g$ and [Fe/H] \citep{Ramirez2009},
studied previously by \citet{Bedell2018}, \citet{Spina2018} and \citet{Botelho2019}. 
The atmospheric parameters are those by \citet{Spina2018}.
The typical errors in $T_{eff}$, $\log g$ and [Fe/H] are 4\,K, 0.012, 0.004\,dex,
respectively, using line-by-line differential spectroscopic analysis relative to 
the Sun by means of equivalent width (EW) measurements of Fe\, I and Fe\,II.
We also adopted the isochronal ages and masses derived by \citet{Spina2018},
who employed the \textbf{q}$^{2}$ code by applying the isochrone method
with the use of the Yonsei-Yale isochrones (as described in \citet{Ramirez2014a, Ramirez2014b}).
This sample spans a wide range in age, i.e. from 500\,Myr up to 8.6\,Gyr
with a typical error of 0.4\,Gyr, which gives us a good understanding of the evolution of the Galactic disc.
We considered the elemental abundances from \citet{Spina2018},
which analysed 12 neutron-capture elements (Sr, Y, Zr, Ba, La, Ce, Pr, Nd, Sm, Eu, Gd, and Dy),
and the abundances of 17 additional elements (C, O, Na, Mg, Al, Si, S, Ca, Sc, Ti, V, Cr, Mn, Co, Ni, Cu and Zn) from \citet{Bedell2018}.
The elemental abundances in both works were also measured based on a line-by-line analysis relative to the Sun.

We have adopted the line-of-sight rotational velocity $V.\sin(i)$ and 
macro-turbulence velocity $V_{\rm macro}$ given in \citet{leonardo} to reproduce the line broadening of our sample of solar twin stars.

The spectra were obtained with the HARPS spectrograph (High Accuracy Radial 
velocity Planet Searcher) on the 3.6m telescope of the ESO (European Southern 
Observatory) La Silla Observatory in Chile \citep{Mayor2003}. The HARPS spectra
cover $\lambda\lambda$3780-6910\,{\AA} with a resolving power (R\,=\,115,000).
Stacking all HARPS spectra of each star resulted in a very high 
signal-to-noise ratio (SNR) of approximately 800\,per\,pixel as measured around 6000\,{\AA},
with a minimum of 300\,per\,pixel and a maximum of 1800\,per\,pixel.
The solar spectrum used in this work is a combination of several exposures of sunlight reflected from the asteroid Vesta,
also observed with HARPS.
Table~\ref{tab_sample} shows the photospheric parameters, broadening velocities,
isochrone age and oxygen abundance of the solar twins sample.

\begin{table*}
\centering
\caption{
Parameters of the 67 solar twins collected from previous published works:
photospheric parameters and isochrone age by \citet{Spina2018},
macro-turbulence and rotation velocities by \citet{leonardo},
and oxygen abundance by \citet{Bedell2018}.
The Sun parameters are added in the first row.
Full table online.
}
\label{tab_sample}
\resizebox{\linewidth}{!}{
\begin{tabular}{rrrrrrrrr}
\hline
Star ID     & $T_{\rm eff}$ & log\,$g$     & [Fe/H]           & $\xi$         & $V_{\rm macro}$ & $V.sin(i)$   & age  & [O/H]\\
            &        (K) &                 & (dex)            & (km.s$^{-1}$) & (km.s$^{-1})$   & (km.s$^{-1}$)& (Gyr) & (dex) \\
\hline
Sun                 & 5777              & 4.440                       &  0.000                      & 1.00                     & 3.20 & 2.04 & 4.56                     & 0.00\\
HIP\,003203 & 5868$\pm$9 & 4.540$\pm$0.016 & -0.050$\pm$0.007 & 1.16$\pm$0.02 & 3.27 & 3.82 & 0.50$\pm$0.30 & -0.139$\pm$0.021\\
HIP\,004909 & 5861$\pm$7 & 4.500$\pm$0.016 &  0.048$\pm$0.006 & 1.11$\pm$0.01 & 3.33 & 4.01 & 0.60$\pm$0.40 & -0.038$\pm$0.017\\
HIP\,006407 & 5775$\pm$7 & 4.505$\pm$0.013 & -0.058$\pm$0.006 & 0.98$\pm$0.01 & 2.96 & 2.30 & 1.90$\pm$0.70 & -0.110$\pm$0.013\\
HIP\,007585 & 5822$\pm$3 & 4.445$\pm$0.008 &  0.083$\pm$0.003 & 1.01$\pm$0.01 & 3.37 & 1.90 & 3.50$\pm$0.40 & 0.054$\pm$0.005\\
--                    & --                      & --                               &  --                              & --                          & --      & --       &  --                       & -- \\
HIP\,118115 & 5798$\pm$4 & 4.275$\pm$0.011 & -0.036$\pm$0.003 & 1.10$\pm$0.01 & 3.55 & 0.89 & 8.00$\pm$0.30 & -0.088$\pm$0.010\\
\hline
\end{tabular}
}
\end{table*}

\section{Determination of C, N and $^{12}$C/$^{13}$C}

The uniform determination of the C abundance was performed from a set of selected absorption lines
of the molecular electronic system A$^{2}\Delta$-X$^{2}\Pi$ of $^{12}$CH (hereafter CH\,A-X, \citet{Jorgensen})
and the N abundance from a set of selected lines of the molecular electronic system
A$^{2}\Sigma^{-}$-X$^{2}\Pi$ of $^{12}$C$^{14}$N (the CN Violet System, hereafter CN\,B-X, \citet{Brooke}).
The homogeneous determination of the $^{12}$C/$^{13}$C isotopic ratio was obtained
by using features from the CH\,A-X system too ($^{12}$CH\,A-X lines mixed with $^{13}$CH\,A-X lines).
We used the C abundance for deriving the N abundance (in the analysis of CN\,B-X lines),
and for extracting the $^{12}$C/$^{13}$C ratio. Although \citet{Bedell2018} have already
determined the C abundance for this same solar twin sample, it was necessary to proceed with
a new determination in order to make a homogeneous chemical analysis,
because \citet{Bedell2018} used equivalent widths of atomic and molecular
lines to derive the carbon abundance (C\,I and CH\,A-X), and in the current work we adopt the spectral
synthesis technique for mensurable and almost isolated lines of CH\,A-X
that were carefully selected based on their sensitivity to the C abundance variation.
We used the same stellar parameters and model atmosphere grid (ATLAS9 by \citet{castelli2004new}),
as presented in \citet{Spina2018}.
We also adopted the same version of the MOOG code, 2014 \citep{moog} that solves the photospheric radiative
transfer with line formation under the LTE (local thermodynamic equilibrium) approximation.
The non-LTE analysis of molecular spectral lines is an unexplored territory in the literature,
but those effects are probably more important in stars with extended atmospheres
\citep{Lambertetal2013}, rather than in our sample of dwarf stars.

Throughout our differential analysis, we have also
adopted the standard parameters for the Sun: $T_{\rm eff}$\,=\,5777\,K, $\log g$\,=\,4.44,
[Fe/H]\,=\,0.00\,dex, and $\xi$\,=\,1.00\,km.s$^{-1}$ \citep[e.g.][]{Cox2000}.
The adopted solar chemical composition is that by \citet{Asplund2009}.
The abundance of oxygen, as derived by \citet{Bedell2018} for the same solar twins sample,
is taken into account along the whole spectral synthesis procedure
for determining the abundances of C and N and the isotopic ratio $^{12}$C/$^{13}$C,
since the molecular dissociative equilibrium is solved by the MOOG code.

The atomic line list has been compiled from the VALD database \citep{Ryabchikova2015},
and we have added the following molecular lines from the Kurucz database \citep{Kurucz}:
$^{12}$C$^{1}$H and $^{13}$C$^{1}$H lines of the A-X system \citep{Jorgensen},
$^{12}$C$^{1}$H and $^{13}$C$^{1}$H lines of the B-X system \citep{Jorgensen},
$^{12}$C$^{12}$C, $^{12}$C$^{13}$C and $^{13}$C$^{13}$C lines of the d-a system \citep{Brooke},
$^{12}$C$^{14}$N, $^{12}$C$^{15}$N and $^{13}$C$^{14}$N lines of the B-X system \citep{Brooke}, and
$^{14}$N$^{1}$H and $^{15}$N$^{1}$H lines of the A-X system \citep{Kurucz2017}.
We have verified the impact of the variation of $^{12}$C/$^{13}$C and $^{14}$N/$^{15}$N,
respectively, on the selected CH\,A-X and CN\,B-X lines,
and we found that those molecular lines are insensitive to change in these isotopic ratios
(the difference in the resulting abundance is always smaller than the abundance error).
The adopted dissociative energies of CH, C$_{2}$, CN and NH are those by \citet{Barklem2016}.

The first step for our differential chemical analysis relative to the Sun was the
selection of (almost) isolated lines of CH\,A-X, CN\,B-X and features of $^{13}$CH\,A-X
($^{13}$CH-$^{12}$CH partially mixed lines). To help us on this search, we adopted the solar
atlas of \citet{Wallace2011}, which identifies various molecular and atomic lines.
We could find, then, the best candidates of CH\,A-X and CN\,B-X non-perturbed lines.
This atlas also allowed us to identify continuum points (or pseudo-continuum in some cases)
for a better flux normalization between observed spectrum and model spectra.
After this step, a detailed investigation is made via spectral
synthesis to address the individual contributions of all species in the region.
Finally, a global $gf$ calibration is made to finely reproduce the HARPS observed solar spectrum
over each selected molecular absorption.
An automated Python code adjusts the $gf$-value under a line-by-line approach for each selected region.
We obtained [X/Fe] near to zero (absolute value always smaller than or equal to 0.005\,dex) for the calibration relative to the Sun
for the derived abundances of C and N in all selected lines,
and $^{12}$C/$^{13}$C between 87.3 and 90.7 for the resulting isotopic ratio in all selected features
(see Fig.~\ref{ch4210}, Fig.~\ref{cn4180} and Fig.~\ref{c4300} in the following sub-sections, in which we show examples).
The next step was to search for continuum points and adequate $\chi^{2}$ windows
for every molecular line/feature, including an inspection of the effective sensitivity to the variation of
the measured parameter (C, N or $^{12}$C/$^{13}$C).
We employ the same procedure to extract the elemental abundance based on the $\chi^{2}$ minimization,
as applied in \citep{Botelho2019}.
The minimizing of $\chi^{2}$ is only computed in a window around the central wavelength of the line/feature,
directly providing the best abundance or isotopic ratio among the synthetic spectra
generated for different values of abundance or isotopic ratio;
$\chi^{2}\,=\,\sum_{i\,=\,1}^{n}(O_{i} - S_{i})^{2}/\sigma(O_{i})^{2}$,
where $O_{i}$ and $S_{i}$ are the flux of the observed and synthetic spectrum, respectively,
$\sigma(O_{i})$ is the error in the observed flux, and $i$ represent the wavelength point.
The observed flux error is estimated as a function of the continuum SNR, that is $\sigma(O_{i})\,=\,O_{i}/$SNR$_{continuum}$.
The $\chi^{2}$ of every spectral synthesis fit is plotted as a function of [X/Fe] for deriving the C and N abundances,
and versus $^{12}$C/$^{13}$C ratio for obtaining this isotopic ratio.
An automated procedure has been used for performing the spectral synthesis fit
in order to measure a final representative abundance or isotopic ratio.

Even though the spectral synthesis has been applied automatically for each molecular line/feature
(CH\,A-X, CN\,B-X and $^{13}$CH\,A-X), we have performed a general visual inspection of all spectral synthesis fits.
The stars HIP\,010303, HIP\,030037, HIP\,038072 and HIP\,083276 were eliminated from our measurements,
because their spectral fits for all CH\,A-X selected lines were unreliable,
not providing a representative abundance of C, and therefore becoming impossible to derive N and $^{12}$C/$^{13}$C.
Perhaps there is a problem with the data reduction in that region for those stars,
as the computed $\chi^{2}$ are very high in comparison with the $\chi^{2}$ distribution of the other stars.

In summary, we were able to measure C, $^{12}$C/$^{13}$C and N
in 63 solar twins from high-resolution high-quality spectra through a
self-consistent and homogenous procedure.

\subsection{Carbon and nitrogen abundances}

In order to derive the carbon abundance,
eleven lines of the (0,0) vibrational band and one of the (1,0) vibrational band of the CH\,A-X system
were selected in the spectral range $\lambda\lambda$4211-4387\,{\AA}.
Table~\ref{tab_CHlines} shows the twelve CH\,A-X lines.
For obtaining the nitrogen abundance, we selected five CN\,B-X lines in 4180-4212\,{\AA},
from the bands (0,0), (1,0) and (2,0), respectively 2, 2 and 1 line.
The five CN\,B-X lines are shown in Table~\ref{tab_CNlines}.
We have also verified that 
the selected lines of $^{12}$CH\,A-X and $^{12}$C$^{14}$N B-X are insensitive
to the variation of the main isotopic ratios of C and N ($^{12}$C/$^{13}$C and $^{14}$N/$^{15}$N). 
Table~\ref{tab_linelistCH} and Table~\ref{tab_linelistCN}, respectively,
present the line list of 2\,{\AA} wide centered at the central wavelength of the selected CH\,A-X and CN\,B-X lines, 
whose $gf$ values have been calibrated to the solar spectrum.

\begin{table*}
\centering
\caption{
The comprehensive list of CH\,A-X lines used in this work for determining the C abundance in solar twin stars.
The line identification (first column) adopts the short molecule notation
and the wavelength of the main molecular electronic transition as a whole number in Angstroms.
The central wavelength of the spectral absorption, assuming a Gaussian profile, is presented in the second column.
The spectral range in which the $\chi^{2}$ is computed and the correspondent number of pixels
are shown in the third and fourth columns, respectively.
The blue and red continuum intervals are presented in sequence,
and, finally, the vibrational band and the number of blended lines to form the molecular absorption are shown in the last columns.
Two CH\,A-X lines represent double absorptions, i.e. two absorptions side by side (CH4248 and CH4278).
}
\label{tab_CHlines}
\resizebox{\linewidth}{!}{
\begin{tabular}{rrrrrrrr}
\hline
line & $\lambda_{central}$ & $\chi^{2}$ window & n. pixels & blue continuum & red continuum & band & n. lines  \\
     & ({\AA})               & ({\AA}) &   & ({\AA}) &  ({\AA}) & ($v'$, $v''$) & \\
\hline
CH4210 & 4210.96 & 4210.83-4211.10 & 27 & 4207.54-4207.60 & 4211.55-4211.62 & (0,0) & 3 \\
CH4212 & 4212.65 & 4212.54-4212.76 & 22 & 4211.56-4211.62 & 4212.94-4213.00 & (0,0) & 3 \\
CH4216 & 4216.60 & 4216.50-4216.70 & 20 & 4214.10-4214.16 & 4218.53-4218.59 & (1,0) & 3 \\
CH4217 & 4217.23 & 4217.11-4217.35 & 24 & 4214.10-4214.16 & 4218.53-4218.59 & (0,0) & 3 \\
CH4218 & 4218.71 & 4218.61-4218.81 & 20 & 4217.90-4217.96 & 4218.93-4218.99 & (0,0) & 3 \\
CH4248 & 4248.72 & 4248.64-4249.01 & 37 & 4246.26-4246.33 & 4251.50-4251.57 & (0,0) & 2 \\
                & 4248.94 &                                  &      &                                  &                                 &          &   1  \\
CH4255 & 4255.25 & 4255.14-4255.36 & 22 & 4253.40-4253.46 & 4256.67-4256.73 & (0,0) & 2 \\
CH4278 & 4278.85 & 4278.77-4279.14 & 37 & 4278.55-4279.59 & 4281.66-4281.70 & (0,0) & 2 \\
                & 4279.06 &                                  &      &                                  &                                 &          &   1  \\
CH4281 & 4281.96 & 4281.84-4282.08 & 24 & 4281.66-4281.70 & 4283.48-4283.54 & (0,0) & 3 \\
CH4288 & 4288.73 & 4288.62-4288.87 & 25 & 4287.20-4287.28 & 4290.55-4290.60 & (0,0) & 4 \\
CH4378 & 4378.25 & 4378.13-4378.37 & 24 & 4377.62-4377.68 & 4379.86-4379.93 & (0,0) & 3 \\
CH4387 & 4387.06 & 4386.94-4387.16 & 22 & 4385.50-4385.58 & 4392.38-4392.45 & (0,0) & 3 \\
\hline
\end{tabular}
}
\end{table*}

\begin{table*}
\centering
\caption{
The comprehensive list of CN\,B-X lines used in this work for determining the N abundance in solar twin stars.
The same notation and layout of Tab.~\ref{tab_CHlines} are adopted.
}
\label{tab_CNlines}
\resizebox{\linewidth}{!}{
\begin{tabular}{rrrrrrrr}
\hline
line & $\lambda_{central}$ & $\chi^{2}$ window & n. pixels & blue continuum & red continuum & band & n. lines  \\
     & ({\AA})               & ({\AA}) &   & ({\AA}) & ({\AA}) & ($v'$, $v''$) & \\
\hline
CN4180 & 4180.02 & 4179.95-4180.08 & 13 & 4179.06-4179.12 & 4181.01-4181.06 & (2,0) & 3 \\
CN4192 & 4192.94 & 4192.84-4193.03 & 19 & 4192.72-4192.78 & 4195.77-4195.85 & (1,0) & 3 \\
CN4193 & 4193.40 & 4193.33-4193.45 & 14 & 4192.72-4192.78 & 4195.77-4195.85 & (1,0) & 3 \\
CN4195 & 4195.95 & 4195.87-4196.01 & 14 & 4195.77-4195.85 & 4203.26-4203.32 & (0,0) & 3 \\
CN4212 & 4212.25 & 4212.15-4212.34 & 19 & 4211.56-4211.63 & 4212.92-4212.98 & (0,0) & 6 \\
\hline
\end{tabular}
}
\end{table*}

\begin{table}
\centering
\caption{
Line lists for the CH\,A-X line regions after the $gf$ calibration to the solar spectrum:
wavelength, species code, excitation potential of transition lower level, $gf$ and species identification.
The species code is the MOOG standard notation, i.e. atomic number(s) before the decimal point (listed in crescent order for molecules)
followed by the ionization level immediately after the decimal point (0: neutral, 1: first ionized, and so on).
Each list covers a region of 2\,{\AA} centered in a given CH\,A-X line.
The first list is for the CH4210 line.
Full tables for all twelve CH\,A-X lines are online.
}
\label{tab_linelistCH}
\begin{tabular}{rrrrl}
\hline
CH4210  &              &        &          &        \\
\hline
wavelength & species code & $\chi_{e}$ & $gf$      & species \\
   ({\AA}) &              &       (eV) &           &         \\
\hline
4209.9650  & 606.0        &     1.559  & 0.316E-03 & C$_{2}$ \\
4209.9710  & 607.0        &     0.258  & 0.135E-05 & CN \\
4209.9760  & 607.0        &     2.348  & 0.682E-04 & CN \\
4209.9760  & 607.0        &     0.998  & 0.941E-05 & CN \\
4209.9870  & 607.0        &     4.001  & 0.289E-03 & CN \\
---        & ---          &     ---    & ---       & --- \\
4211.9670  & 606.0        &     1.574  & 0.257E$+$02 & C$_{2}$ \\
\hline
\end{tabular}
\end{table}

\begin{table}
\centering
\caption{
Line lists for the CN\,B-X line regions after the $gf$ calibration to the solar spectrum.
The same notation of Tab.~\ref{tab_linelistCH} is adopted.
Each list covers a region of 2\,{\AA} centered in a given CN\,B-X line.
The first list is for the CN4180 line.
Full tables for all five CN\,B-X lines are online.
}
\label{tab_linelistCN}
\begin{tabular}{rrrrl}
\hline
CN4180 &              &        &          &        \\
\hline
wavelength & species code & $\chi_{e}$ & $gf$      & species \\
   ({\AA}) &              &       (eV) &           &         \\
\hline
4179.0220  & 607.0        &     3.822 & 0.226E-02  & CN \\
4179.0270  & 606.0        &     1.515 & 0.211E-09  & C$_{2}$ \\
4179.0360  &   24.0        &     3.849 & 0.486E-03  & Cr\,I \\
4179.0380  & 107.0        &     1.139 & 0.110E-06  & NH \\
4179.0390  & 607.0        &     3.518 & 0.211E-02  & CN \\
---        & ---          &     ---   & ---        & --- \\
4181.0140  & 606.0        &     1.538 & 0.469E-08  & C$_{2}$ \\
\hline
\end{tabular}
\end{table}

\begin{figure*}
\includegraphics[scale=0.270]{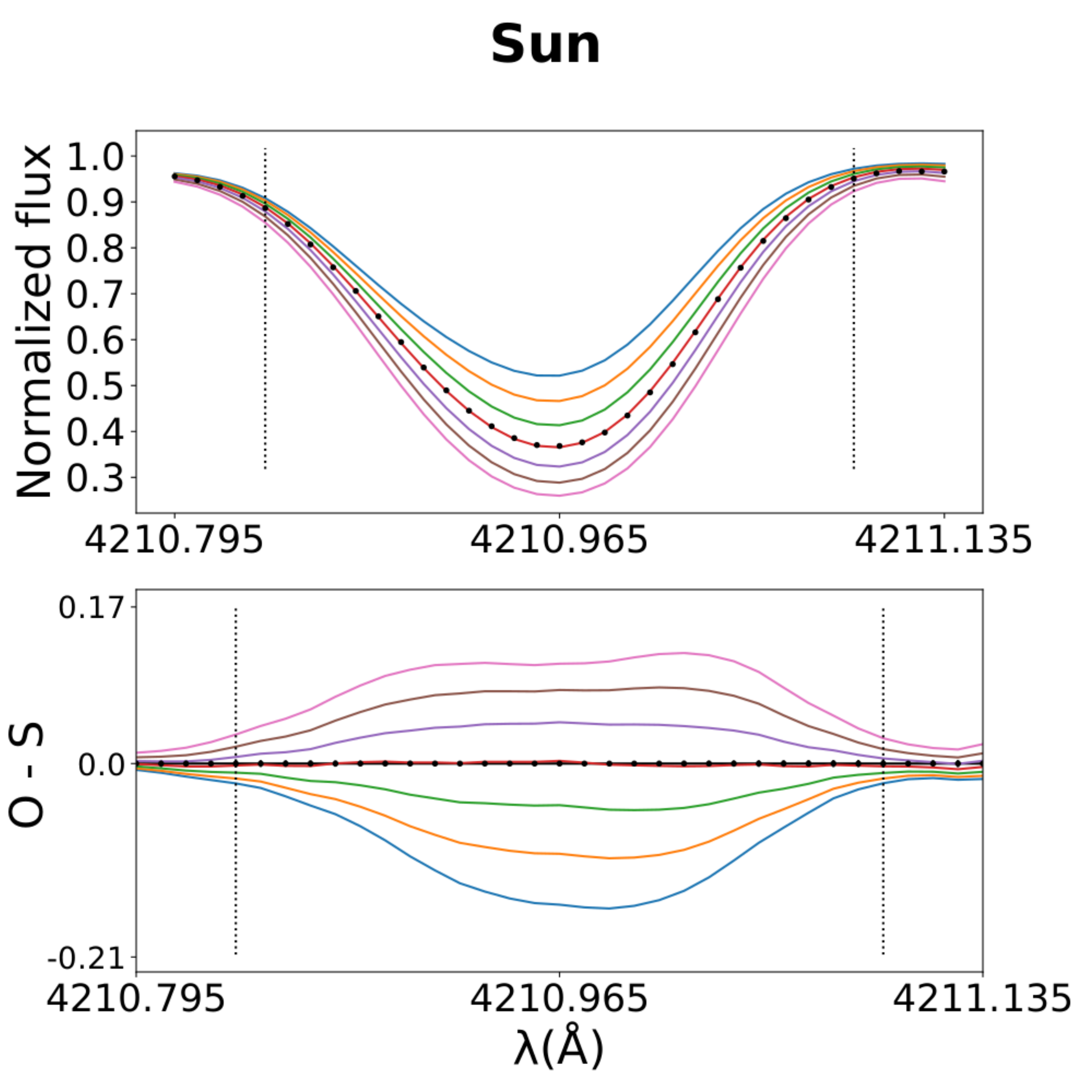}
\includegraphics[scale=0.246]{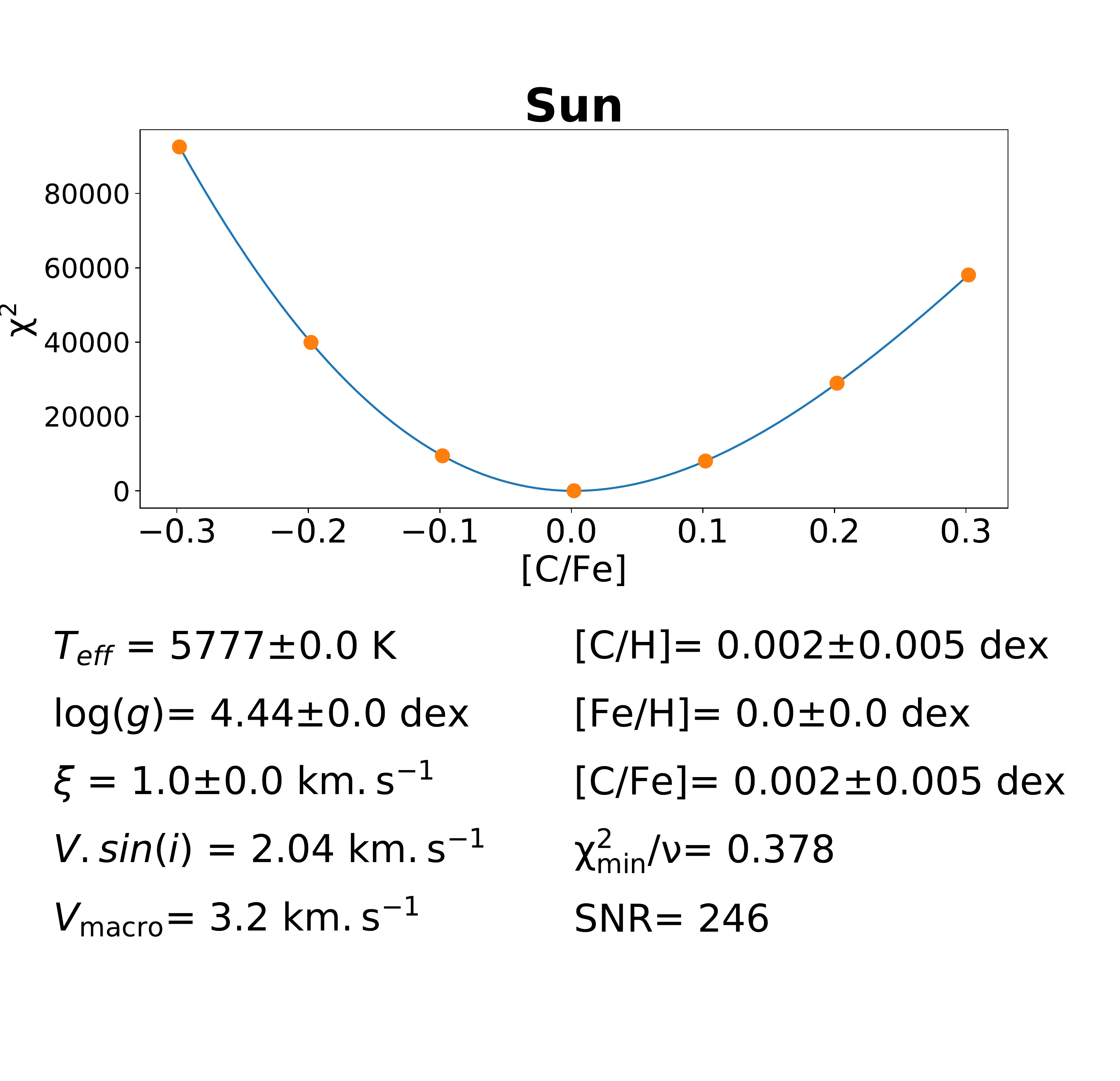}
\includegraphics[scale=0.255]{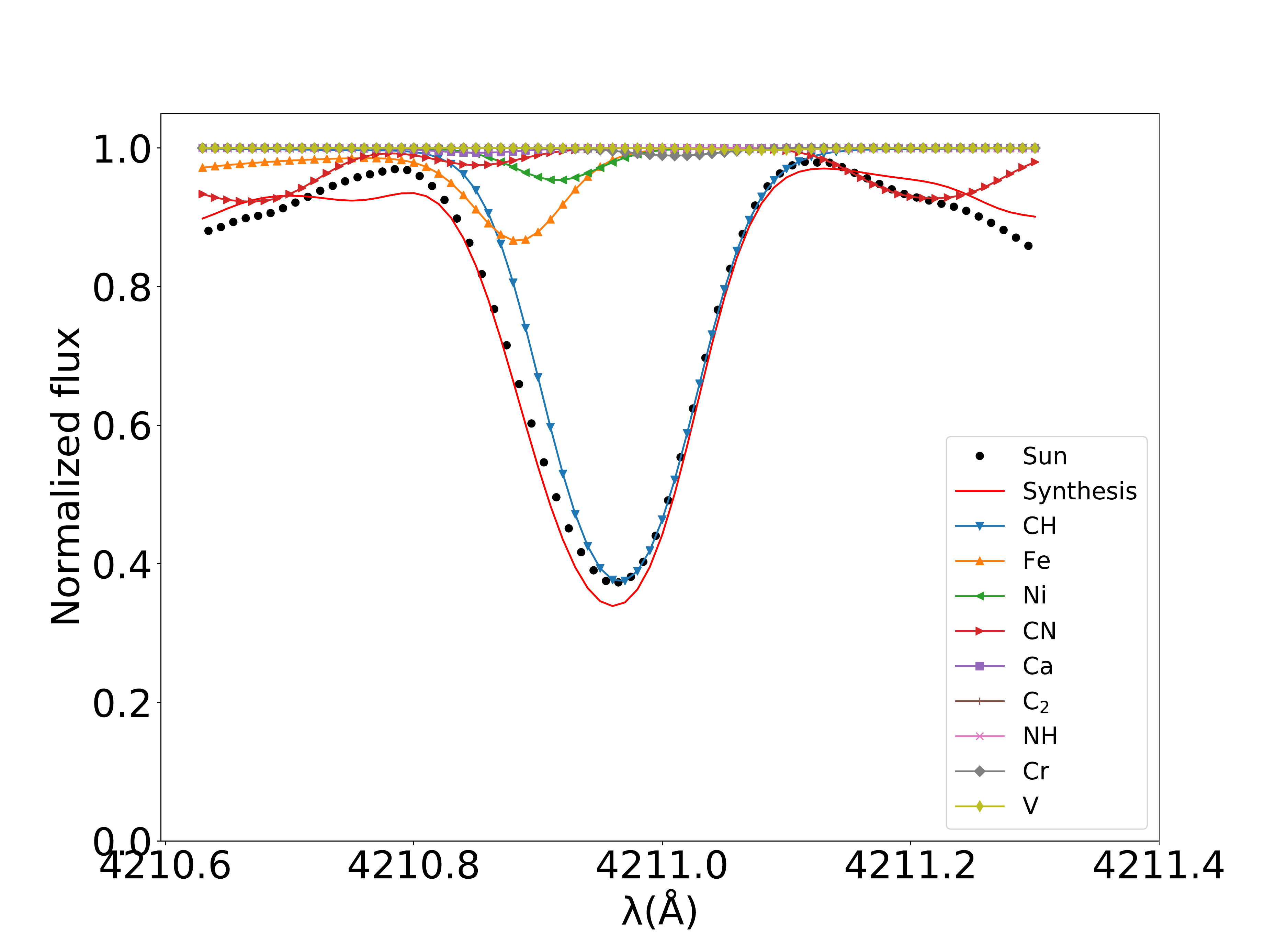}
\caption{
Example of spectral synthesis calibration to the Sun of a CH\,A-X line (CH4210),
showing the individual contributions from different species in the line profile (bottom panel),
spectral comparisons between the observed spectrum and seven synthetic spectra (top-left panel),
and the $\chi^{2}$ graph as a function of [X/Fe] (top-right panel).
}
\label{ch4210}
\end{figure*}

\begin{figure*}
\includegraphics[scale=0.270]{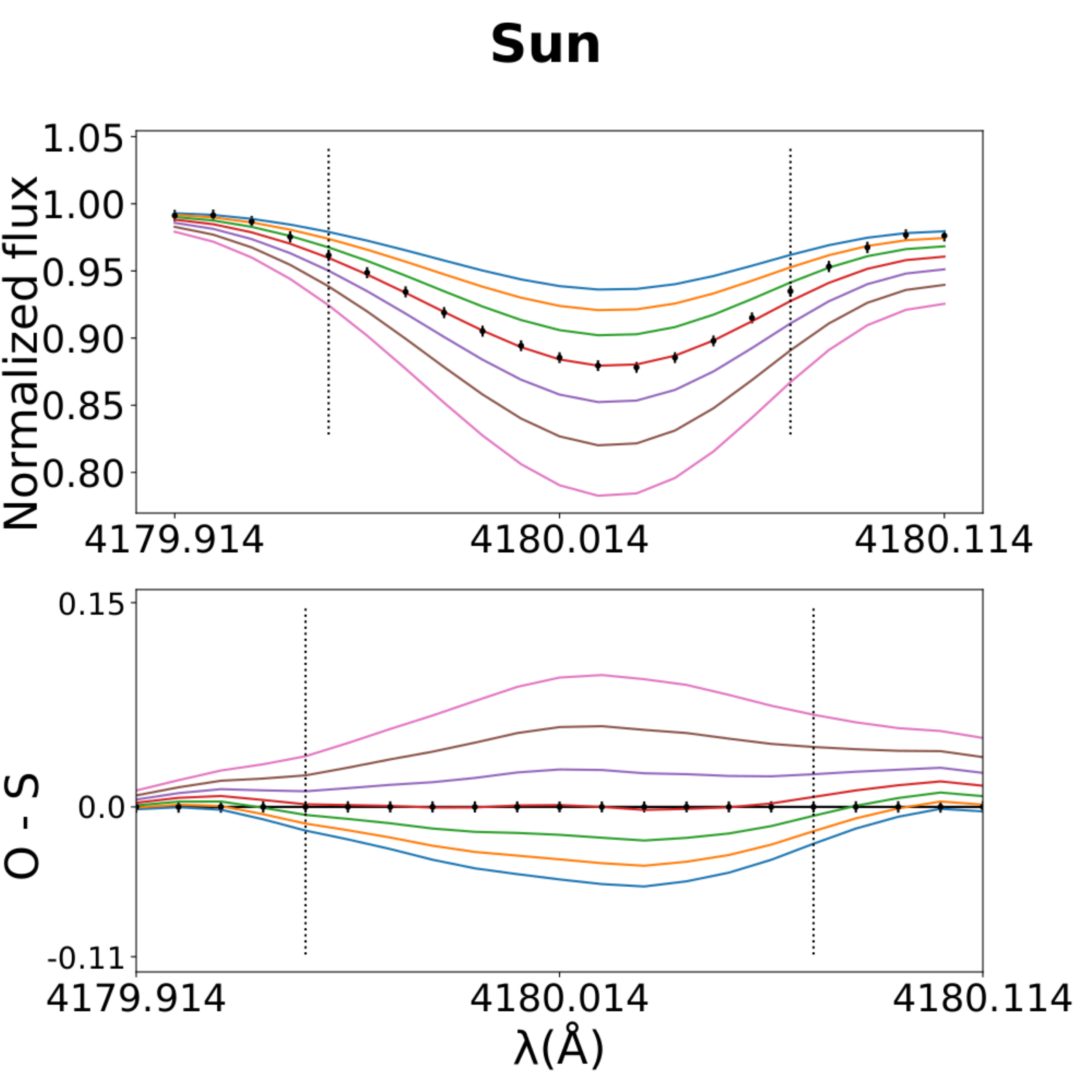}
\includegraphics[scale=0.246]{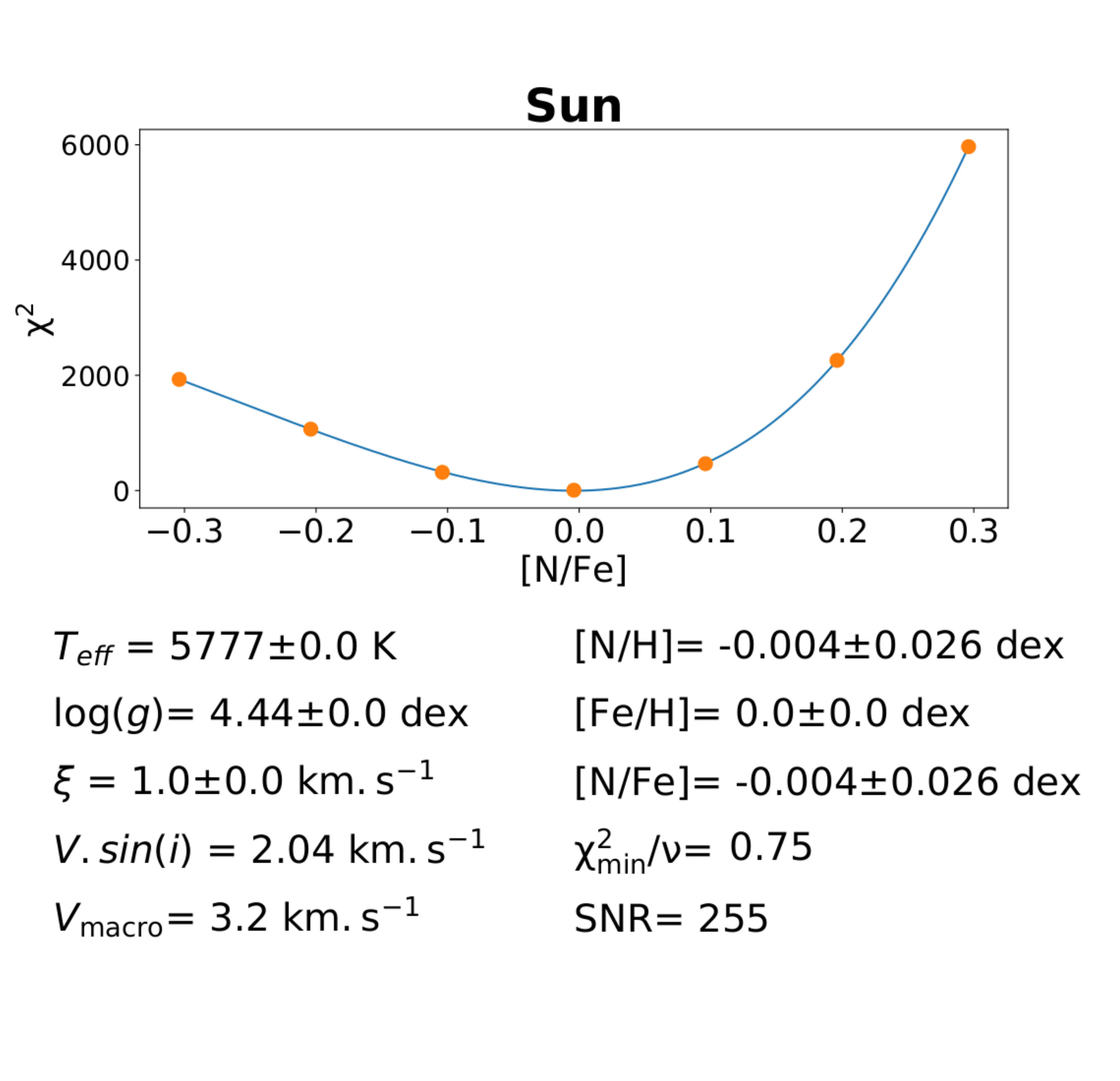}
\includegraphics[scale=0.255]{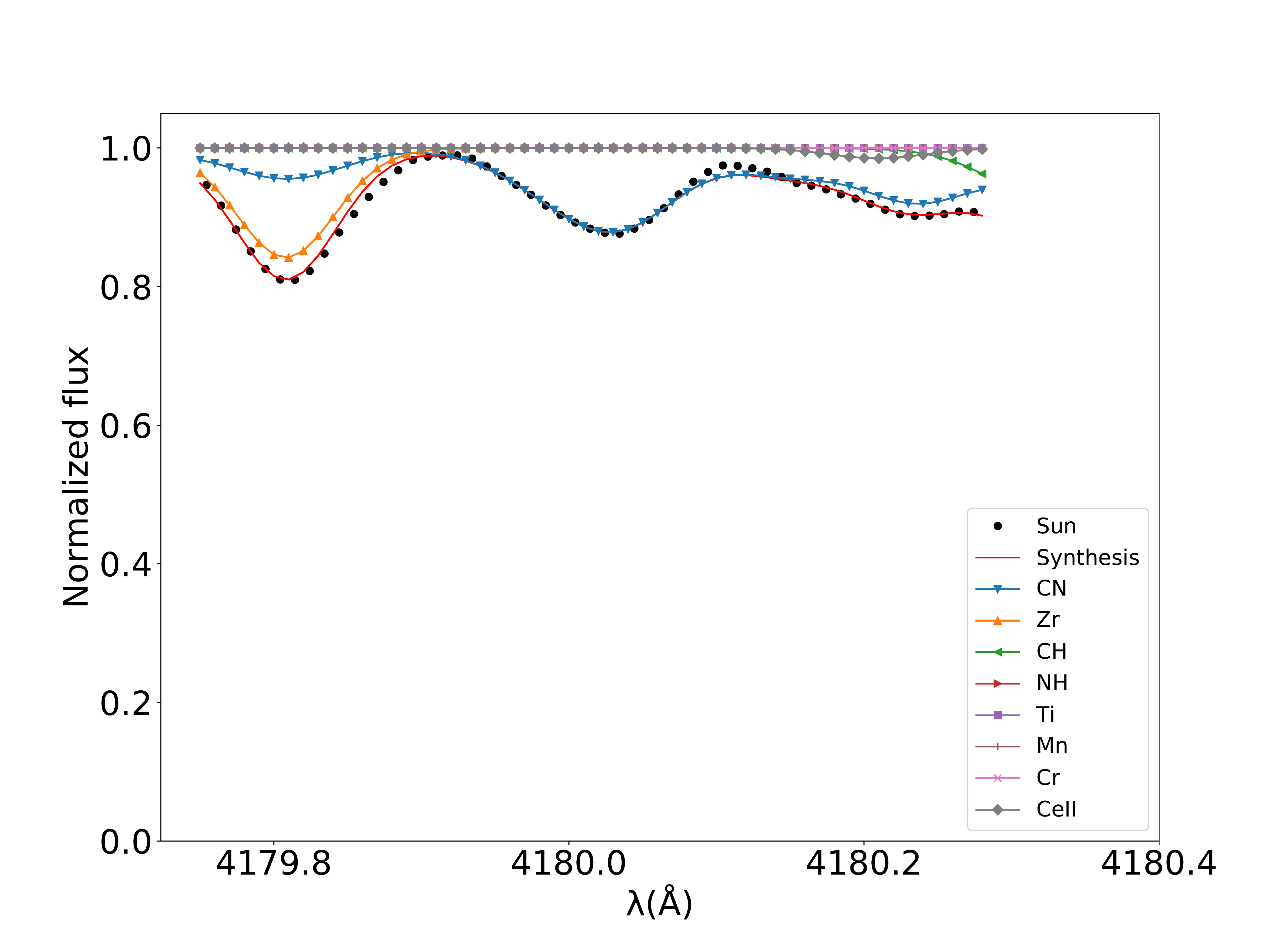}
\caption{
Example of spectral synthesis calibration to the Sun of a CN\,B-X line (CN4180).
The same plots of Fig.~\ref{ch4210} are adopted.
}
\label{cn4180}
\end{figure*}

For performing the spectral synthesis fit of every selected line of CH\,A-X and CN\,B-X,
seven synthetic spectra are computed with uniform step of 0.10\,dex in [X/Fe] (X: C and N respectively)
that are resampled in wavelength to that sampling of the observed spectrum.
After that, the continuum level of the observed spectrum needs to be fitted to the continuum level of each synthetic spectrum;
the continuum correction is multiplicative and derived by using both blue and red continuum ranges
(see Tab.~\ref{tab_CHlines} and Tab.~\ref{tab_CNlines}).
For deriving the resulting elemental abundance from each molecular line,
the $\chi^{2}$ between the model and observed spectra is computed in the line window.
The $\chi^{2}$ window covers from 20 up to 37 pixels for the CH\,A-X lines and from 13 up to 19 pixels for the CN\,B-X lines.
The C abundance is derived first (as a simple mean) and then the N abundance is obtained (simple average too), after fixing the C abundance.
Both C and N abundances were used to measure of the $^{12}$C/$^{13}$C isotopic ratio,
making the overall chemical analysis self-consistent.
Figure~\ref{ch4210} and Figure~\ref{cn4180} are examples of spectral synthesis calibration to the solar spectrum,
from which the C and N measurements are derived from, respectively, a CH\,A-X line and a CN\,B-X line,
showing diagnostic plots for the individual contributions from different species in the line profile
and for the spectral synthesis itself as well as the graph of $\chi^{2}$ as a function of [X/Fe],
whose minimum directly provides the resulting elemental abundance. 
None resulting abundance from every CH\,A-X and CN\,B-X line was excluded by a 3$\sigma$ clipping criterium.

The $\chi^{2}$ minimization procedure directly provides a good estimation for the error in [X/Fe].
The abundance error is derived when $\chi^{2}$ increases by $\nu$ degrees of freedom
from the minimum $\chi^{2}$ value along the polynomial fitting curve $\chi^{2}$ versus [X/Fe],
such that $\nu$ is the number of pixels in the $\chi^{2}$ window less a single free parameter (the elemental abundance itself).
The uncertainties in abundance due to the photospheric parameter errors
plus the impact of the continuum level adjustment, considering the influence of the flux noise,
have been added in quadrature to that error from the spectral synthesis itself ($\chi^{2}$ estimation).
For estimating the final error in the N abundance, the whole carbon abundance error is also taken into account.
We have also estimated the impact of the oxygen abundance error
in the abundances of carbon of nitrogen, and we found that the influence is negligible
(i.e. the difference in the resulting abundance is always smaller than 1\,milidex).
The error in [X/H] is computed in quadrature with the error in [Fe/H].
The global error in [X/Fe] derived from each molecular line is roughly dependent on the local SNR
(measured in the spectral region of the molecular lines).
The SNR changes from 250\,per\,pixel up to 600\,per\,pixel for the CH\,A-X line regions,
and from 200\,per\,pixel up to 500\,per\,pixel for the CN\,B-X line regions.

Concerning the C abundance derivation by \citet{Bedell2018} for the same solar twins sample,
we have found a systematic difference in the [C/H] scale of about -0.04\,dex
between our determinations and
their average [C/H] abundance ratios based on EW measurements from C\,I and CH\,A-X lines.
However, this small discrepancy is lower than the data dispersion
(e.g. a standard $lsq$ linear fit provides a $rms$ of 0.08\,dex).

The final abundances in C and N ([X/Fe] scale) for the 63 analyzed solar twins are computed as simple averages
of the measured abundances from the selected individual molecular lines.
The errors in the final [C/Fe] and [N/Fe] are estimated as simple means of their individual values.
The error in [C/Fe] varies from 0.004 up to 0.011\,dex with an average value around 0.006\,dex
(the error in [C/H] changes from 0.004 up to 0.013\,dex).
The error in [N/Fe] varies from 0.015 up to 0.057\,dex, whose average value lies around 0.030\,dex
(the error in [N/H] changes from 0.016 up to 0.057\,dex).

\subsection{$^{12}$C/$^{13}$C isotopic ratio}

In order to self-consistently derive the $^{12}$C/$^{13}$C isotopic ratio,
our previous measurements of the C and N concentrations were adopted.
We found six $^{13}$CH-$^{12}$CH\,A-X features around the G band of CH at $\lambda$4300\,{\AA}
(all containing lines of the (0,0) vibrational band of the CH\,A-X system)
that are sensitive to the variation of the $^{12}$C/$^{13}$C ratio.
Table~\ref{tab_CClines} shows these CH\,A-X selected features (in fact, $^{13}$CH\,A-X lines mixed with $^{12}$CH\,A-X lines).
Table~\ref{tab_linelistCC} presents the line list covering 2\,{\AA}
centered at the central wavelength of the $^{13}$CH-$^{12}$CH\,A-X selected features, 
whose $gf$ values have been calibrated to the solar spectrum.

\begin{table*}
\centering
\caption{
The comprehensive list of $^{13}$CH-$^{12}$CH\,A-X features used in this work
for determining the $^{12}$C/$^{13}$C isotopic ratio in solar twin stars.
The same notation and layout of Tab.~\ref{tab_CHlines} are adopted.
}
\label{tab_CClines}
\resizebox{\linewidth}{!}{
\begin{tabular}{rrrrrrrr}
\hline
feature & $\lambda_{central}$ & $\chi^{2}$ window & n. pixels & blue continuum & red continuum & band & n. lines  \\
     & ({\AA})               & ({\AA}) &   & ({\AA}) & ({\AA}) & ($v'$, $v''$) & \\
\hline
$^{13}$CH4297  & 4297.10 & 4297.03-4297.17 & 14 & 4292.83-4292.88 & 4304.93-4305.01 & (1,0) &  6 \\
$^{13}$CH4298  & 4298.14 & 4298.08-4298.21 & 13 & 4292.83-4292.88 & 4304.93-4305.01 & (1,0) &  3 \\
$^{13}$CH4299A & 4299.42 & 4299.34-4299.51 & 17 & 4292.83-4292.88 & 4304.93-4305.01 & (1,0) &  7 \\
$^{13}$CH4299B & 4299.74 & 4299.65-4299.83 & 18 & 4292.83-4292.88 & 4304.93-4305.01 & (1,0) &  8 \\
$^{13}$CH4300  & 4300.67 & 4300.58-4300.76 & 18 & 4292.83-4292.88 & 4304.93-4305.01 & (1,0) &  5 \\
$^{13}$CH4303  & 4303.32 & 4303.24-4303.40 & 16 & 4292.83-4292.88 & 4304.93-4305.01 & (1,0) &  3 \\
\hline
\end{tabular}
}
\end{table*}

\begin{figure*}
\includegraphics[scale=0.270]{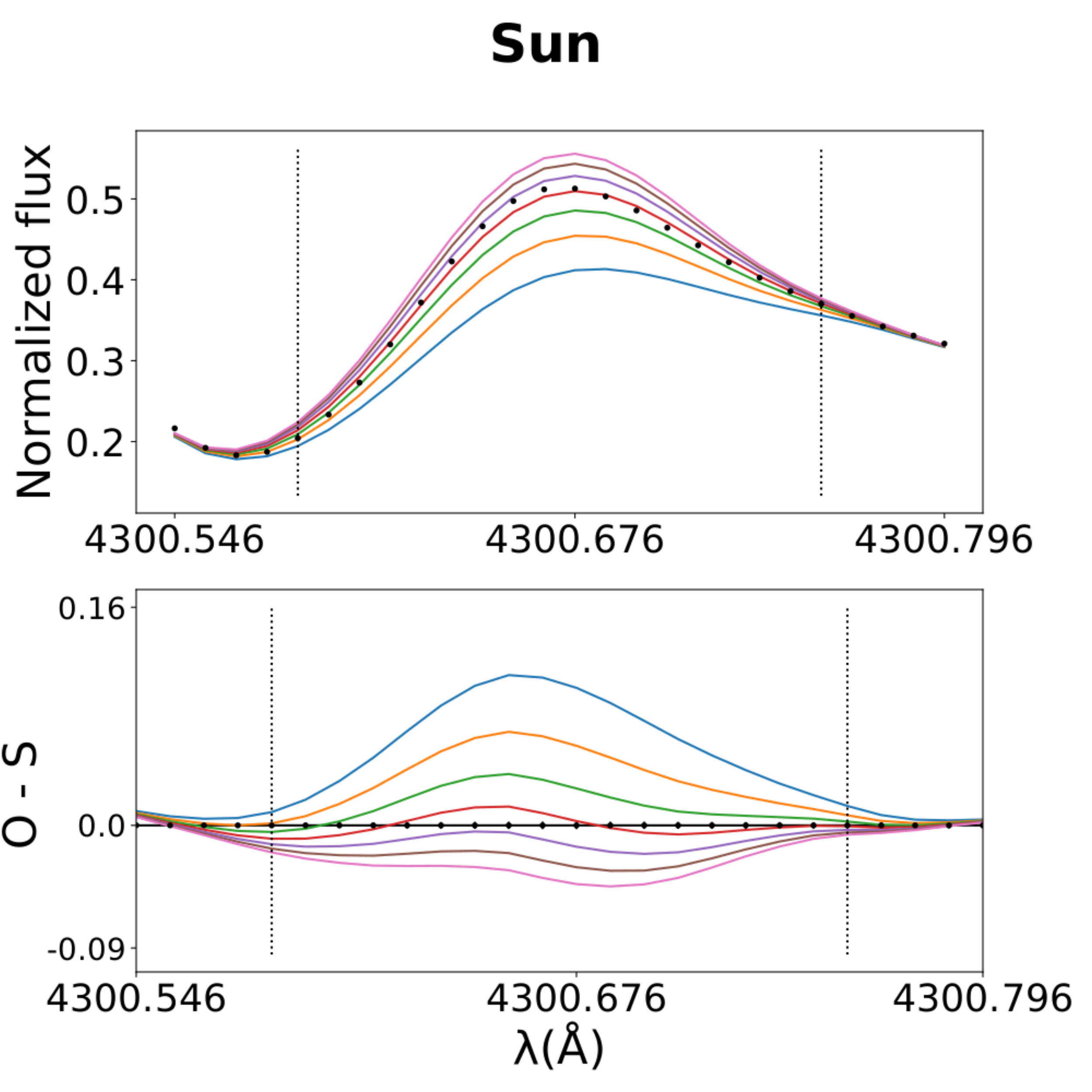}
\includegraphics[scale=0.246]{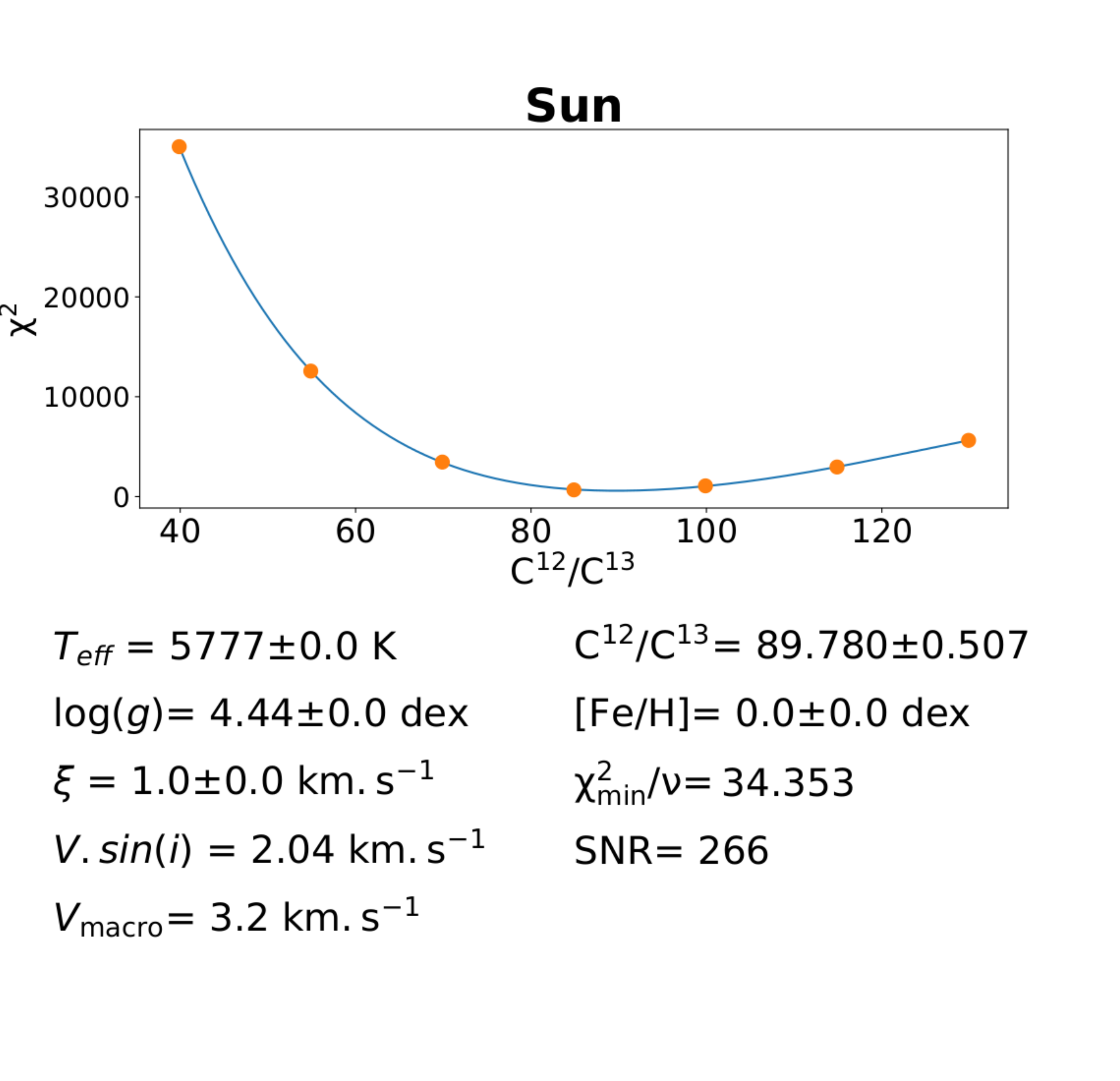}
\includegraphics[scale=0.255]{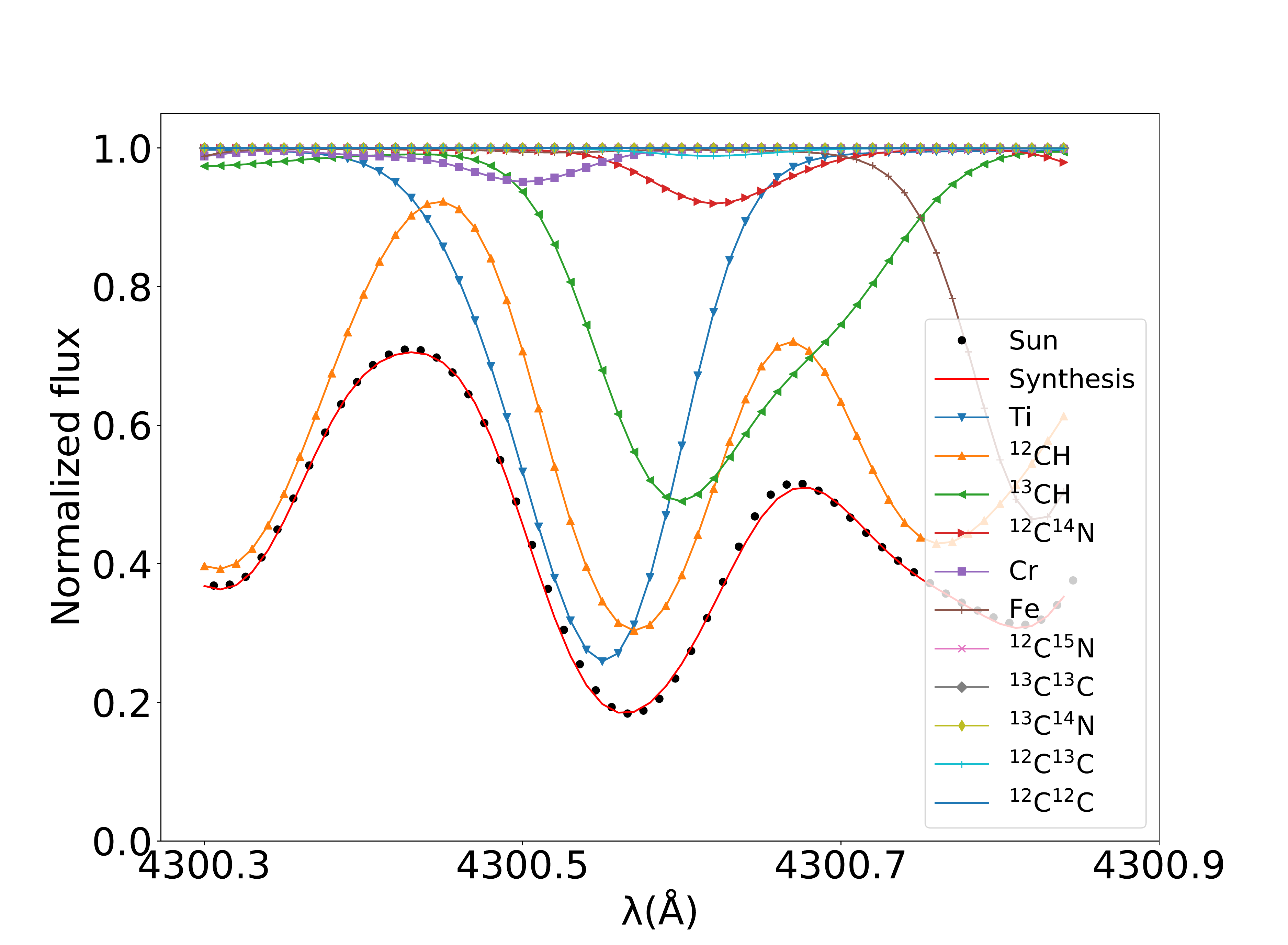}
\caption{
Example of spectral synthesis calibration to the Sun of a $^{13}$CH-$^{12}$CH\,A-X selected feature ($^{13}$CH4300).
The same plots of Fig.~\ref{ch4210} are adopted, except in the top-right panel,
which shows $\chi^{2}$ as a function of $^{12}$C/$^{13}$C instead.
}
\label{c4300}
\end{figure*}

Analogously to the measurements of the C and N abundances,
seven synthetic spectra are computed to derive the $^{12}$C/$^{13}$C isotopic ratio assuming a uniform step of 15 in this free parameter.
The model spectra are resampled in wavelength to the sampling of the observed spectrum.
The observed continuum level is also individually fitted
to the continuum level of each synthetic spectrum (see Tab.~\ref{tab_CClines}).
For deriving the resulting isotopic ratio from each $^{13}$CH-$^{12}$CH\,A-X selected feature,
the $\chi^{2}$ between the model and observed spectra is also computed along the feature window (covering from 13 up to 18\,pixels).
Both C and N abundances were used to measure of the $^{12}$C/$^{13}$C isotopic ratio, making the overall chemical analysis self-consistent.
Along the procedure of spectral synthesis, whilst the isotopic ratio $^{14}$N/$^{15}$N is fixed to the solar value,
the ratio $^{12}$C/$^{13}$C to be measured starts from the solar value.
We have adopted the values suggested by \citet{Asplund2009} for the Sun
($^{14}$C/$^{15}$C\,=\,89.4$\pm$\,0.2 and $^{14}$N/$^{15}$N\,=\,435$\pm$\,57).
Figure~\ref{c4300} is an example of spectral synthesis calibration to the solar spectrum,
from which $^{12}$C/$^{13}$C is derived by using a $^{13}$CH-$^{12}$CH\,A-X selected feature.
This figure shows diagnostic plots for the individual contributions from different species in the line profile and
for the spectral synthesis itself as well as the graph of $\chi^{2}$ versus $^{12}$C/$^{13}$C,
in which the isotopic ratio is recovered from the $\chi^{2}$ minimum value.

\begin{table}
\centering
\caption{
Line lists for the $^{13}$CH-$^{12}$CH\,A-X feature regions after the $gf$ calibration to the solar spectrum.
The same notation of Tab.~\ref{tab_linelistCH} is adopted,
also including the mass numbers of atoms (listed in crescent order) in the species code in the case of molecular lines,
but excluding the species identification (last column in the previous tables).
Each list covers a region of 2\,{\AA} centered in a given $^{13}$CH-$^{12}$CH\,A-X feature.
The first list is for the $^{13}$CH4297 feature.
Full tables for all six $^{13}$CH-$^{12}$CH\,A-X features are online.
}
\label{tab_linelistCC}
\begin{tabular}{rrrr}
\hline
$^{13}$CH4297 &                  &                   &       \\
\hline
wavelength & species code & $\chi_{e}$ & $gf$ \\
   ({\AA})       &                          &       (eV)      &           \\
\hline
 4296.1010  & 606.01313    &     2.145 & 0.731E-01 \\
 4296.1030  & 606.01313    &     1.329 & 0.637E-01 \\
 4296.1060  & 606.01213    &     2.043 & 0.111E-01 \\
 4296.1060  & 607.01314    &     2.349 & 0.302E-04 \\
 4296.1080  & 106.00112    &     1.247 & 0.550E-02 \\
          ---        & ---                    &     ---        & ---                \\
 4298.1000  & 607.01214    &     0.566 & 0.259E-03 \\
\hline
\end{tabular}
\end{table}

By adopting the 3$\sigma$ clipping criterium over the measurements of the resulting $^{12}$C/$^{13}$C isotopic ratios for every sample star,
four $^{13}$CH-$^{12}$CH\,A-X features have been used for 7 stars,
five features for other 7 stars,
and all of them have been useful for the 49 remaining stars.

The estimation of the error in $^{12}$C/$^{13}$C is similar to that performed for the C and N abundances,
i.e. the $\chi^{2}$ minimization approach objectively gives the error due to the spectral synthesis procedure itself.
The basic error in $^{12}$C/$^{13}$C is estimated from the variation of $\chi^{2}$ in ($\nu$\,-\,1) unities
along the polynomial fitting curve $\chi^{2}$-$^{12}$C/$^{13}$C, where $\nu$ is the number of pixels in the $\chi^{2}$ window.
The error due to the spectral synthesis is added in quadrature
with the propagated photospheric parameter errors, the uncertainties in the C abundance
and the errors from the impact of the continuum level adjustment.
We have also estimated the impact of the error in the oxygen abundance over the error of the isotopic ratio,
and we found that the influence is negligible.
We have verified that the global error in $^{12}$C/$^{13}$C derived from each molecular feature
is roughly dependent on the local SNR (measured in the correspondent spectral regions).
The SNR changes from 170\,per\,pixel up to 400\,per\,pixel for the $^{13}$CH-$^{12}$CH\,A-X feature regions.

Like the C and N abundances, the final $^{12}$C/$^{13}$C isotopic ratios for the 63 analyzed solar twins
are computed as simple averages of the individual measurements from the $^{13}$CH-$^{12}$CH\,A-X features,
and their errors as simple means of the individual errors.
The error in $^{12}$C/$^{13}$C varies from 1.9 up to 10.5, having 4.3 as an average value.

\section{Analysis of the results}

We have measured C, $^{12}$C/$^{13}$C and N in 63 solar twins from HARPS high-resolution high-quality spectra
through a self-consistent, homogeneous and automated procedure.
Seven old $\alpha$-enhanced solar twins (HIP\,014501, HIP\,028066, HIP\,030476, HIP\,065708, HIP\,073241, HIP\,074432 and HIP\,115577)
and the single solar twin anomalously rich in $s$-process elements HIP\,064150 were excluded from the further analysis,
in order to be consistent with the analyses carried out
by \citet{Bedell2018} and \citet{Spina2018} on other elements for the same solar twins sample,
representing a Galaxy's homogeneous population (i.e. the local thin disc \citep{Ramirezetal2012}).
Therefore, the remaining 55 thin disc solar twins are investigated
for studying correlations of [C/Fe], [N/Fe], $^{12}$C/$^{13}$C, [C/N], [C/O] and [N/O] with [Fe/H] and age,
as well as [C/N], [C/O] and [N/O]  versus [O/H].

The nearby $\alpha$-enhanced stars are in average older than the ordinary thin disc stars \citep{Adibekyanetal2011, Haywoodetal2013}.
The $\alpha$-rich stars, which have [<$\alpha$/Fe>] higher in about 0.1\,dex,
are split into two Galaxy's disc different stellar populations,
carrying on two distinct dynamical histories, in contrast with the local thin disc stars.
Whilst the metal-poor $\alpha$-rich stars actually belong to the thick disc based on their kinematics and orbital parameters,
the metal-rich $\alpha$-rich stars exhibit nearly circular orbits close to the Galactic plane (similarly to the thin disc stars).
Those seven old $\alpha$-enhanced solar twins could be either thick disc 'metal-rich' stars \citep{Bedell2018}
or might have likely come from inner disc regions,
as speculated by \citet{Adibekyanetal2011} for the metal-rich $\alpha$-rich stars.

\begin{figure}
\center
\includegraphics[scale=0.300]{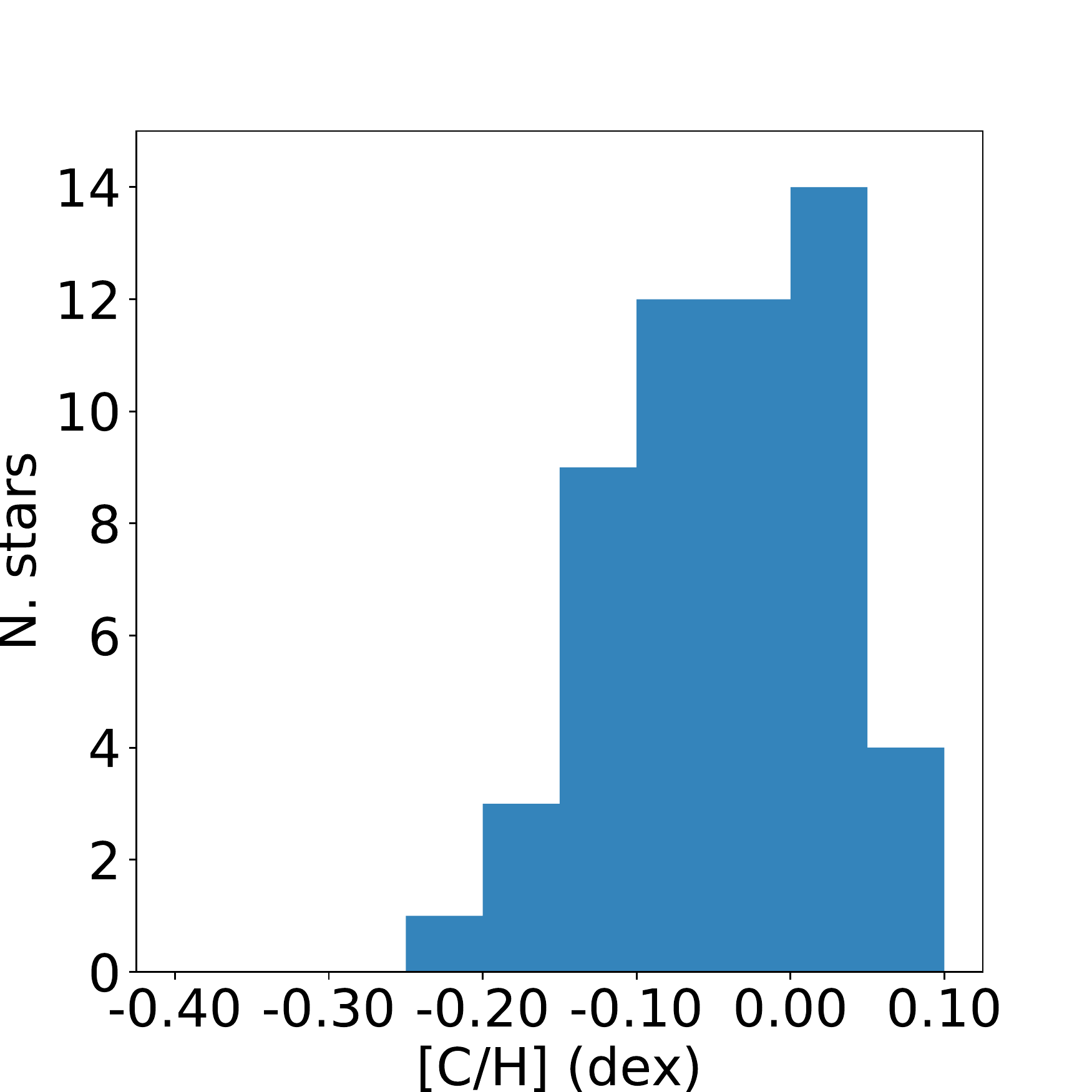}
\includegraphics[scale=0.300]{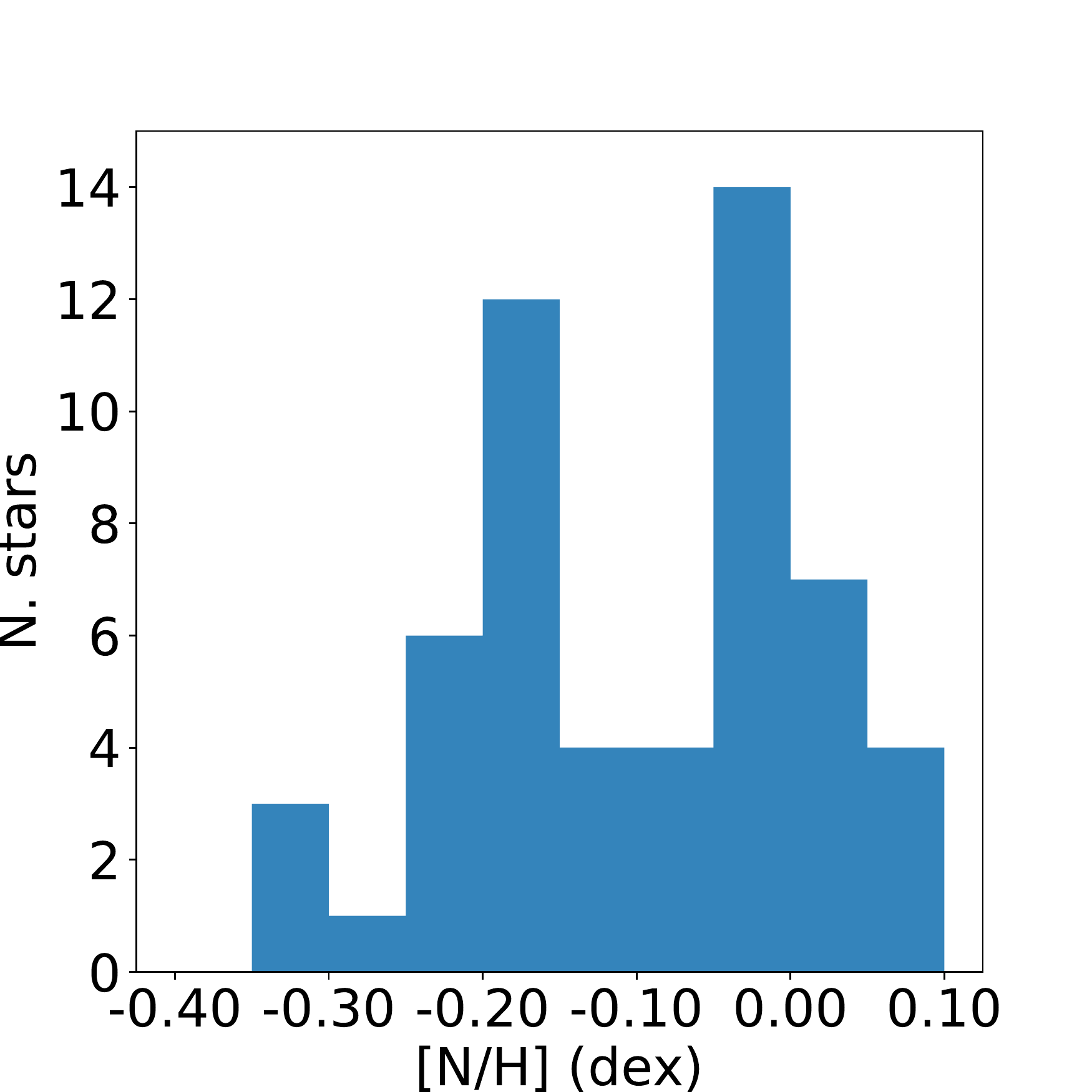}
\includegraphics[scale=0.300]{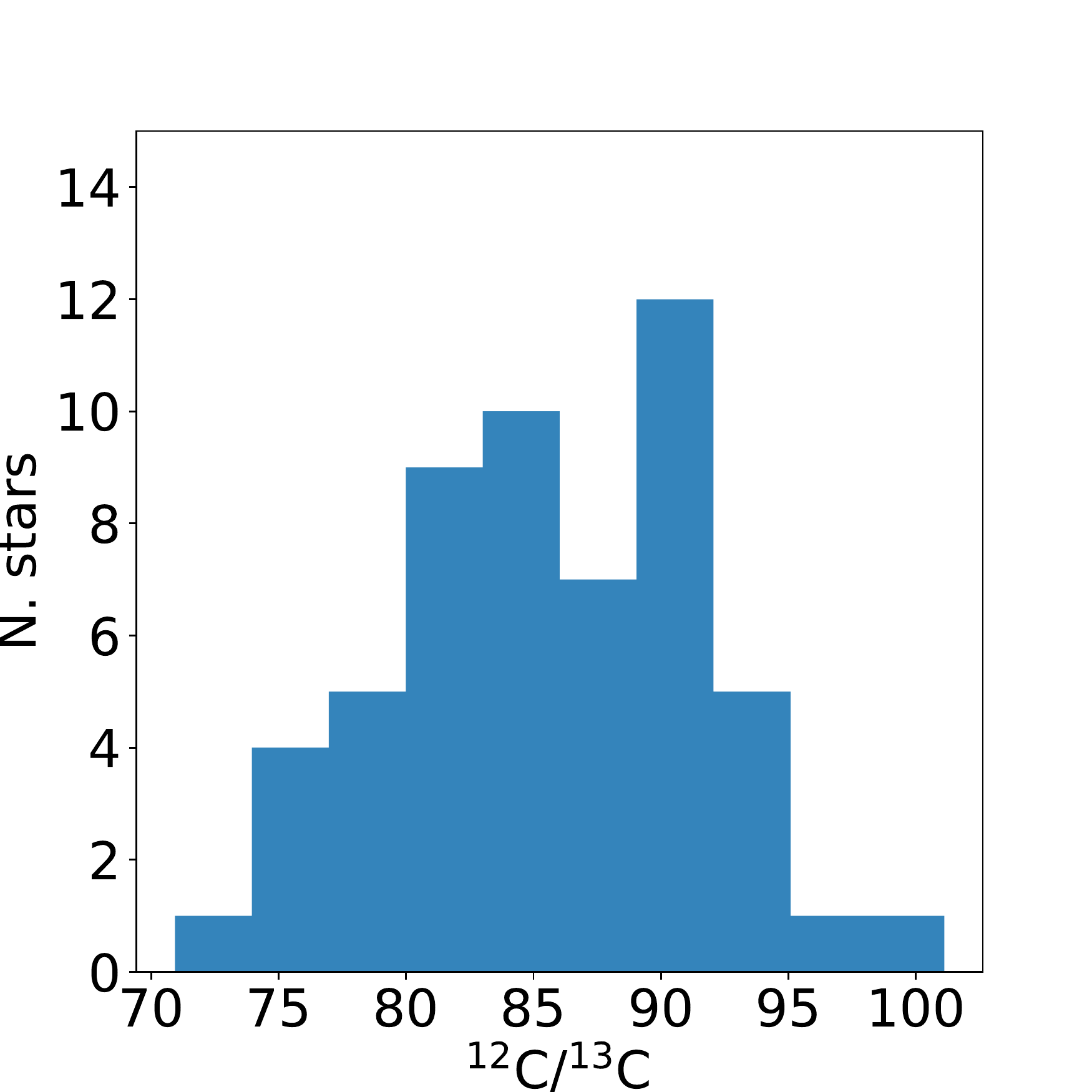}
\caption{
Distributions of [C/H], [N/H] and $^{12}$C/$^{13}$C among the 55 thin disc solar twins.
}
\label{hist}
\end{figure}

The determination of the C abundance, N abundance and $^{12}$C/$^{13}$C ratio reached, respectively,
100, 100 and 80\,per\,cent of completeness in terms of all selected molecular lines under the 3$\sigma$-clipping criterium.
Table~\ref{results} compiles our measurements for the 63 solar twins.
Figure~\ref{hist} shows the distributions of [C/H], [N/H] and $^{12}$C/$^{13}$C among 55 thin disc solar twins.

Regarding the 55 thin disc solar twins,
[C/Fe] changes from -0.129 up to 0.042\,dex and [C/H] from -0.207 up to 0.080\,dex.
Their average values, respectively, are about -0.040\,dex ($\sigma$\,=\,0.033\,dex),
and -0.042\,dex ($\sigma$\,=\,0.070\,dex).
The mean errors of [C/Fe] and [C/H] are, respectively, 0.006 and 0.007\,dex;
their respective variations are: 0.004-0.011\,dex and 0.004-0.013\,dex. 
[N/Fe] changes from -0.313 up to 0.023\,dex and [N/H] from -0.310 up to 0.087\,dex.
Their average values, respectively, are about -0.094\,dex ($\sigma$\,=\,0.075\,dex),
and -0.096\,dex ($\sigma$\,=\,0.108\,dex).
The average errors in [N/Fe] and [N/H] are the same, i.e. around 0.030\,dex.
The standard deviations of the [N/Fe] and [N/H] errors are, respectively, 0.016 and 0.057\,dex.

Also regarding the 55 thin disc solar twins,
the isotopic ratio $^{12}$C/$^{13}$C varies from 70.9 up to 101.1,
presenting an average value around 85.8 ($\sigma$\,=\,6.2).
The error in $^{12}$C/$^{13}$C varies from 1.9 up to 10.5, having 4.3 as an average value (like for all 63 solar twins).

We have investigated [C/Fe], [N/Fe], $^{12}$C/$^{13}$C, [C/N], [C/O] and [N/O] as a function of [Fe/H] and isochrone stellar age,
and also [C/N], [C/O] and [N/O] as a function of [O/H].
Only $^{12}$C/$^{13}$C is analysed as a function of [C/Fe] and [N/Fe].
We adopt linear dependence for all relations, based on those 55 thin disc solar twins,
whose results are discussed in the following subsections.
We have used the Kapteyn kmpfit package\footnote{https://www.astro.rug.nl/software/kapteyn/index.html}
to do all fits to the data.
The KMPFIT code performs a robust linear fit $y$ versus $x$,
which minimizes the orthogonal distance of the overall data points to the fitting curve,
taking into account the errors in both variables $x$ and $y$ under a variance weighting approach.
The Sun data are not included in all linear fits.
Table~\ref{value-adj} compiles the results of all computed linear fits for
[C/Fe], [N/Fe], $^{12}$C/$^{13}$C, [C/N], [C/O] and [N/O] as a function of [Fe/H] and age;
Table~\ref{value-adj-razao} for $^{12}$C/$^{13}$C as a function of [C/Fe] and [N/Fe]; and
Table~\ref{value-adj-oh} for [C/N], [C/O] and [N/O] as a function of [O/H].

\begin{table*}
\centering
\caption{
Carbon and nitrogen abundances and $^{12}$C/$^{13}$C ratio measured in this work relatively to the Sun for 63 solar twins.
The Sun data are included in the first row. Full table online.
}
\label{results}
\begin{tabular}{rrrrrr}
\hline
Star ID   & {[}C/H{]}        & {[}C/Fe{]}       & {[}N/H{]}        & {[}N/Fe{]}       & $^{12}$C/$^{13}$C \\ 
          & (dex)            & (dex)            & (dex)            & (dex)            &                   \\
\hline
Sun                &   0.002$\pm$0.001 &  0.002$\pm$0.001 &  0.002$\pm$0.018 &   0.002$\pm$0.018 & 88.7$\pm$0.5 \\
HIP\,003203 & -0.129$\pm$0.012 & -0.079$\pm$0.010 & -0.301$\pm$0.056 & -0.251$\pm$0.056 & 84.9$\pm$9.8 \\
HIP\,004909 & -0.018$\pm$0.011 & -0.066$\pm$0.009 & -0.122$\pm$0.042 & -0.170$\pm$0.042 & 89.3$\pm$8.5 \\
HIP\,006407 & -0.092$\pm$0.012 & -0.034$\pm$0.010 & -0.180$\pm$0.045 & -0.122$\pm$0.045 & 87.9$\pm$7.5 \\
HIP\,007585 &  0.040$\pm$0.006 & -0.043$\pm$0.005 & -0.035$\pm$0.024 & -0.118$\pm$0.024 & 87.0$\pm$3.2 \\
--       &     --             &    --              &     --            &    --              &      --        \\
HIP\,118115 & -0.054$\pm$0.007 & -0.018$\pm$0.006 & -0.196$\pm$0.030 & -0.160$\pm$0.030 & 85.1$\pm$4.0 \\
\hline
\end{tabular}
\end{table*}

\subsection{[C/Fe] and [N/Fe] versus [Fe/H] and age}

\begin{figure*}
\includegraphics[scale=0.266]{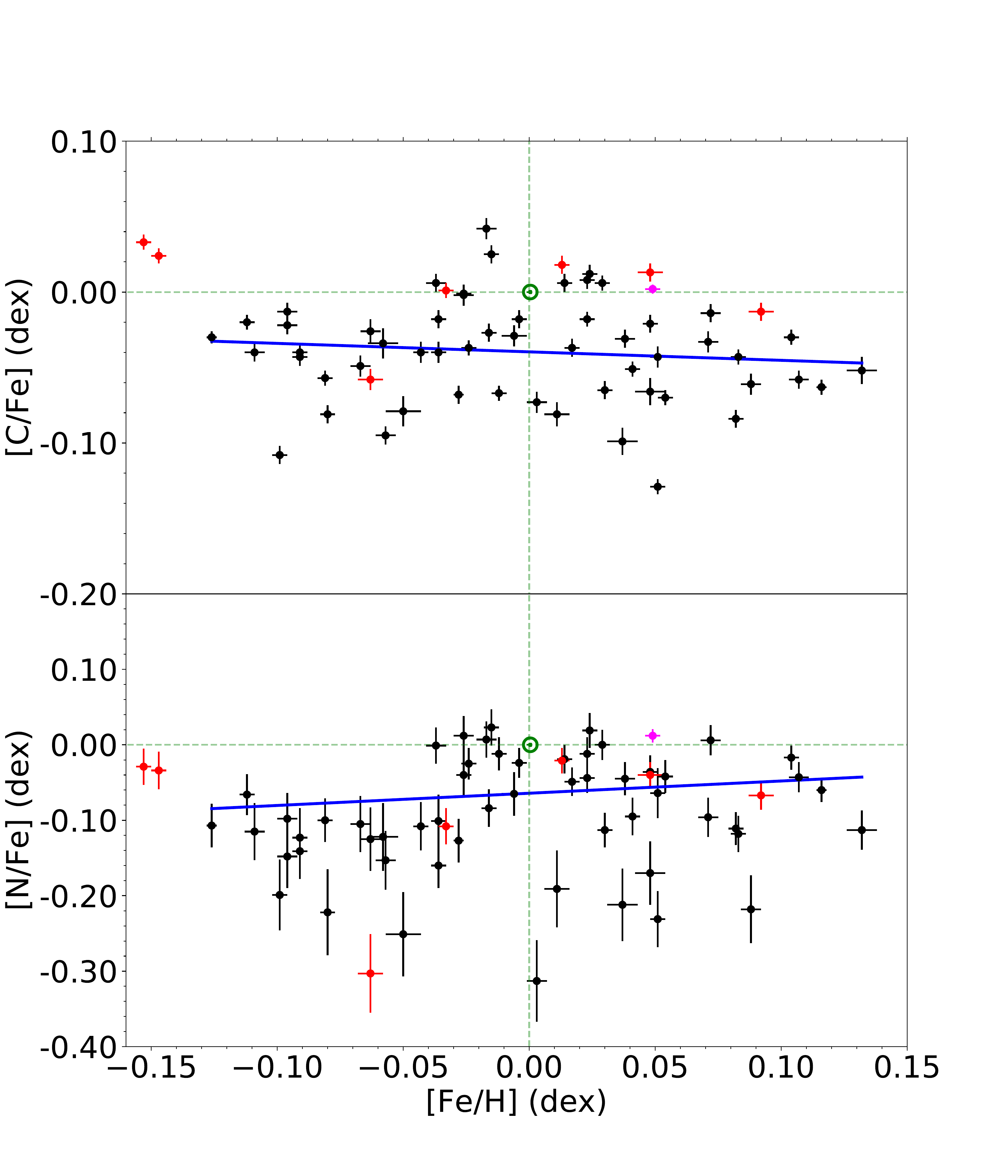}
\includegraphics[scale=0.372]{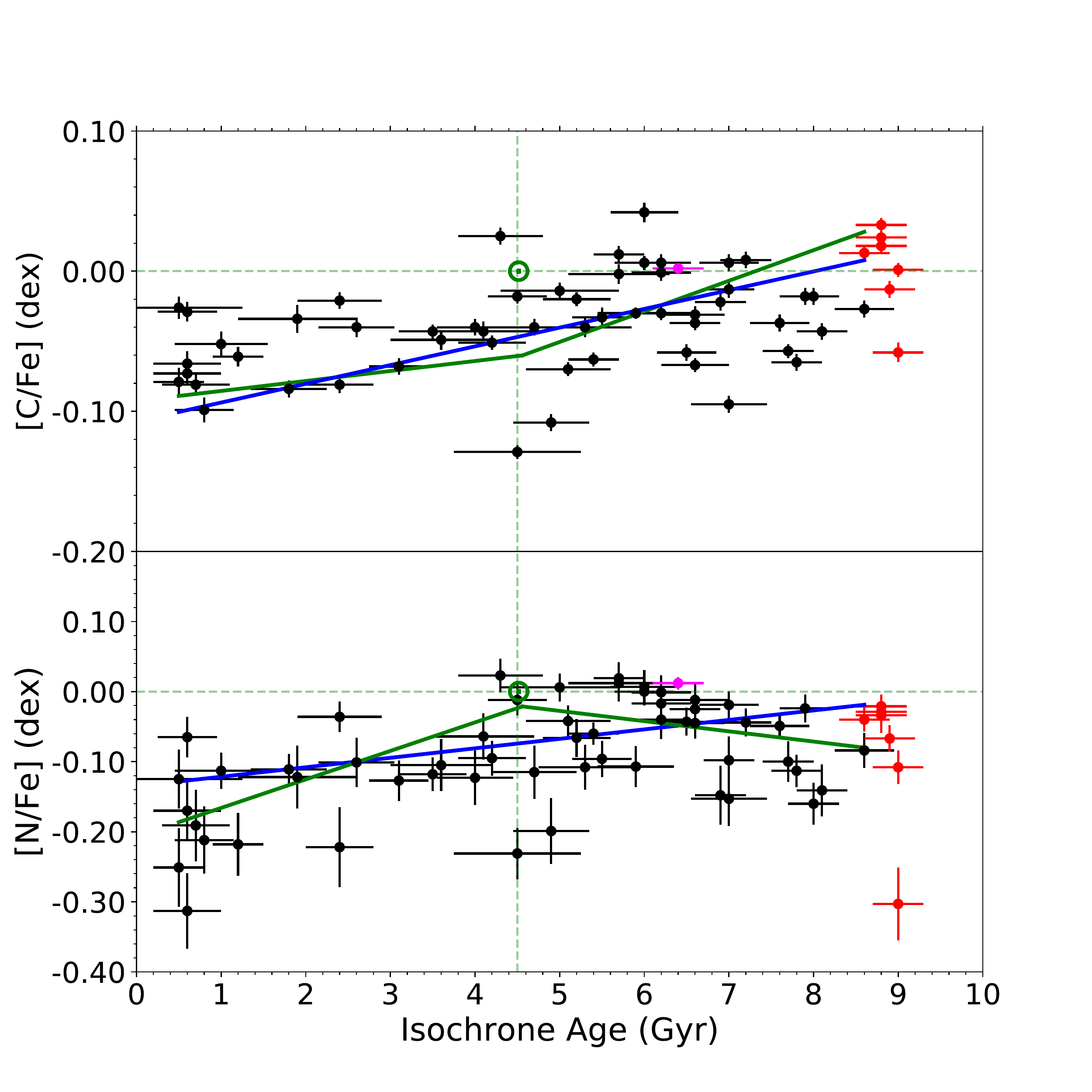}
\caption{
[C/Fe] and [N/Fe] as a function of [Fe/H] and isochrone stellar age for the 55 thin disc solar twins:
the global linear fit is shown in every plot (blue solid line),
and the broken linear fit is only presented in the [X/Fe]-age plots (green solid line),
splitting the sample into stars with ages up to the Sun's age and stars older than the Sun
(the break is fixed to the Sun's age). 
The red points represent the seven old $\alpha$-rich solar twins
and the magenta point represents the star rich in $s$-process elements HIP\,064150.
These stars do not take part in the fits as well as the Sun,
whose data are also plotted as reference (green symbol).
The results of each linear fit are shown in Table~\ref{value-adj}.
}
\label{cfe-nfe-feh-age}
\end{figure*}

Concerning the relations between [C,N/Fe] and [Fe/H]
in the small range of metallicity of solar twins (Fig.~\ref{cfe-nfe-feh-age}),
we have found that:
{\bf (i)} [C/Fe] is anti-correlated with [Fe/H] with a slope of -0.056$\pm$0.012
(the relative error of the negative slope is about 21\,per\,cent); and
{\bf (ii)} there is a tentative correlation between [N/Fe] and [Fe/H]
(slope relative error of 35\,per\,cent or just 2.8$\sigma$ of significance).

The Sun seems slightly above,
when compared against the fits of [C/Fe] and [N/Fe] versus [Fe/H] of the solar twins
(within about 1\,$rms$ of deviation in average from the linear relations).
This result is expected indeed, because \citet{Melendez2009} found,
as confirmed afterwards by other works \citep{Ramirez2009, Ramirez2010, Nissen2015, Bedell2018},
that the Sun is slightly deficient in refractory elements relative to volatile ones in comparison with most of solar twins.
Being C and N volatile elements and Fe a refractory element,
thus the Sun is slightly enhanced in C and N relative to Fe.
The 50\,per\,cent equilibrium condensation temperatures of C, N, O and Fe,
for the solar nebula and the Solar System composition taking into account the total pressure at 1\,AU, 
are respectively 40, 123, 180 and 1334 \,K \citep{Lodders2003}.

We can see relatively good agreements by comparing our derived relations [C/Fe]-[Fe/H] and [N/Fe]-[Fe/H]
with those by \citet{DaSilva2015} done over a more extended range in [Fe/H] than solar twins, for a sample of 120 thin disc FGK dwarfs. 
Restricting the results from \citet{DaSilva2015} to the small metallicity range of solar twins,
there would be a negative trend between [C/Fe] and [Fe/H] and no trend between [N/Fe] and [Fe/H],
with the Sun data also slightly above the overall data set.
Unfortunately, \citet{DaSilva2015} did not publish the slopes
for the anti-correlation and correlation [C/Fe] vs. [Fe/H] found respectively for stars with [Fe/H] below and above zero,
neither for the positive correlation [N/Fe] vs. [Fe/H] found along the [Fe/H] whole range.
\citet{Suarez-Andres2017} recently presented linear fits between [C/Fe] and [Fe/H]
for a huge sample of 1110 solar-type stars.
They also derived [C/Fe]-[Fe/H] fits for stars with [Fe/H] below and above the solar value.
We cannot do the same unfortunately, because the [Fe/H] amplitude of our solar twins sample is much smaller.
On the other hand, we are able to split our sample in terms of stellar age due to a wide coverage in this parameter
(see the following paragraphs and last subsection in the case of the carbon abundance ratios).
The \citet{Suarez-Andres2017}'s sample was split into two groups of stars with planets of different masses
and a large comparison sample of stars without detected planets.
Additionally, they performed linear fits between [C/H] and [Fe/H] for these subsamples
over the whole wide metallicity range (-1.4\,$\leq$\,[Fe/H]\,$\leq$\,+0.6\,dex).
For the stellar sample without planets, \citet{Suarez-Andres2017} found a slope of +0.970$\pm$0.008
that agrees within 1$\sigma$ with the slope derived by us for this relation (+0.989$\pm$0.015).
Note that the solar twins sample of the current work has only two known planet-host stars so far.
On the other hand, \citet{Nissen2015} studied the chemical content of a sample of 21 solar twins like us
(just 3 of them holding known planets).
However, \citet{Nissen2015} did not provide any fit for the [C/Fe]-[Fe/H] relation.
By inspecting the results for carbon, there seems to be a negative trend between [C/Fe] and [Fe/H],
as derived by us for a larger solar twins sample.
\citet{Suarez-Andres2016} obtained, for a restricted sample of 32 solar-type stars without known planets,
no correlation between [N/Fe] and [Fe/H] in the range -0.45\,$\leq$\,[Fe/H]\,$\leq$\,+0.55\,dex
(the slope that they found has a large error, i.e. +0.040$\pm$0.070).
The whole sample of \citet{Suarez-Andres2016} has 74 stars, 42 of which holding detected planets until the publication date.
The slope for the tentative correlation [N/Fe]-[Fe/H] derived by us for solar twins also agrees within 1$\sigma$
with the result from \citet{Suarez-Andres2016}.

The linear fits of [C,N/Fe] as a function of the isochrone stellar age
over the wide range of ages of our solar twins sample (Fig.~\ref{cfe-nfe-feh-age}) provide the following results:
{\bf (i)} [C/Fe] is positively well correlated with the stellar age (slope of +0.013$\pm$0.001\,dex.Gyr$^{-1}$
with a relative error of 8\,per\,cent, i.e. 13$\sigma$ of significance); and
{\bf (ii)} [N/Fe] is also positively well correlated with the stellar age
(slope of +0.013$\pm$0.002\,dex.Gyr$^{-1}$ with a relative error of about 15\,per\,cent).
The Sun again seems slightly above,
when compared with the derived fits of both [C/Fe] and [N/Fe] versus age of the solar twins 
(1.2\,$rms$ of deviation in average from the linear relations),
confirming the Sun is slightly deficient in refractory elements relative to volatile ones
in comparison with solar twins \citep{Melendez2009, Nissen2015, Bedell2018}.

The analysis of the abundance ratios as a function of the stellar age
is extended with linear fits that fix a break point at the Sun's age, whose value is very well stablished
(the fit results are added in Tab.~\ref{value-adj}).
The sample of 55 thin disc solar twins is split into
stars with ages up to the Sun's age (23 {\lq\lq}young{\rq\rq} solar twins)
and stars with ages greater than the solar age (32 {\lq\lq}old{\rq\rq} solar twins).
Analogously to the global single linear fits, the alternative broken linear fits  (applied with the Python.Scipy ODR function)
perform the minimization of the orthogonal distances of the data points to the fitting function
taking into account the errors in both variables.
By fixing the abscissa of the break point, the fitting computation is simplified (four free parameters instead of five)
and two linear functions are derived with a connection/transition at the solar age.
Each broken linear fit is shown associated with the global fit in the same plot
(right panel in Fig.~\ref{cfe-nfe-feh-age}, bottom panel in Fig.~\ref{razao-feh-age} plus Fig.~\ref{artigo_cn_co_no_age}).
For instance, \citet{DaSilva2015} applied independent linear fits of [X/Fe] versus isochrone stellar age
for a sample of FGK dwarfs also considering ages below and above the Sun's age
(X: C, N, O, Na, Mg, Si, Ca, Ti, V, Mn, Fe, Ni, Cu and Ba).

Whilst [C/Fe] is found to have a tentative correlation with age for the {\lq\lq}young{\rq\rq} solar twins,
it is correlated with age for the {\lq\lq}old{\rq\rq} solar twins,
such that the broken linear fit matches well the overall behaviour of [C/Fe] based on the global linear fit (Fig.~\ref{cfe-nfe-feh-age}).
The broken linear fit exhibits overall effective abundance ratio dispersion comparable to the dispersion of the global single fit.
We estimate an effective $rms$ of 0.039\,dex over the whole age scale against 0.037\,dex of the global fit
($rms_{\rm effective}$\,$\approx$\,sqrt(($rms_{\rm young\,twins}^{2}$\,+\,$rms_{\rm old\,twins}^{2}$)/2)).
The broken linear fit [N/Fe]-age gives a stronger correlation with age for the {\lq\lq}young{\rq\rq} solar twins in comparison with the global fit
(slope of +0.041($\pm$0.008)\,dex.Gyr$^{-1}$),
and a negative trend for the {\lq\lq}old{\rq\rq} solar twins (see Fig.~\ref{cfe-nfe-feh-age} and Tab.~\ref{value-adj}).
However, there is an overall decrease of [N/Fe] with time from the older solar twins towards the younger solar twins,
like seen in the global single fit.
The effective dispersion of the broken fit over the whole age scale is also equivalent to the dispersion of the global single fit
(overall $rms$ of 0.067\,dex against 0.070\,dex of the global fit).

If we could consider our solar twins relations [C,N/Fe]-age as roughly representative for the local thin disc evolution,
rude estimates for the current [C/Fe] and [N/Fe] ratios in the local ISM
would be respectively -0.11 and -0.13\,dex (as suggested by the global fits' linear coefficients).
In addition to these estimates, the global fits also suggest that [C/Fe] and [N/Fe]
would be respectively around 0.00 and -0.02\,dex for 8.6\,Gyr ago in the local ISM,
which is close to {\lq\lq}the formation epoch{\rq\rq} of the thin disc.
The [C/Fe]-age broken fit for the solar twins younger and older than the Sun
also provides the same rude estimates for the current and ancient values of [C/Fe] in the local ISM.
In fact, the precision of absolute values for both abundance ratios estimated for the past and the present has less statistical significance
in comparison with their overall distribution and variation in time.
At least, we are confident to predict that both [C/Fe] and [N/Fe] in the local ISM are certainly under-solar nowadays
as well as they were close to the solar ratios in average 8.6\,Gyr ago.

Therefore, [C/Fe] and [N/Fe] have decreased
along the evolution of the Galactic thin disc in the solar neighbourhood during nearly 9\,Gyr.
This is likely due to a higher relative production of iron in the local disc (by basically SN-Ia in binary systems)
in comparison with the nucleosynthesis of carbon and nitrogen
(by single low-mass/intermediate-mass AGB stars plus metal-rich massive stars too).
Whilst the carbon-to-iron ratio decreases in time and with the iron abundance,
the nitrogen-to-iron ratio decreases in time but seems to increase with the iron abundance.
Consequently, it is complicated to reach a conclusive statement
about the impact of the ratio AGB-stars/SN-Ia or even the frequency of isolated stars relative to binary stars,
to consistently explain the lowering of the carbon-to-iron and nitrogen-to-iron ratios over time.

We have found a few GCE models \citep{Kobayashi2011, Sahijpal2013}
that partially agree with our estimates for the [C/Fe] and [N/Fe] abundance ratios at the solar neighbourhood.
The GCE models with AGB yields by \citet{Kobayashi2011} predict [N/Fe] around 0.0\,dex
for {\lq\lq}the formation epoch{\rq\rq} of the thin disc
(see Fig.\,13 in this publication considering [Fe/H]\,=\,-1.5\,dex as the minimum metallicity for the thin disc stars),
predicting a $^{12}$C/$^{13}$C isotopic ratio around 80 in the current time (see Figs.\,17-19 in this publication).
The models case A by \citet{Sahijpal2013} predict [C/Fe] around -0.1\,dex in the current time (see Fig.\,5 in this paper),
and the grid model\,\#30 of the case A given by the same work specifically predicts [C/Fe] around the solar value
for {\lq\lq}the formation epoch{\rq\rq} of the thin disc
(also see Fig.\,5 in this publication, assuming an initial [Fe/H]\,=\,-1.5\,dex).

Besides these comparisons against theoretical predictions,
all the GCE models by \citet{Romano2019} predict an anti-correlation between [C/Fe] and [Fe/H] around the solar ratios, like we observe,
and three of them show a positive correlation between [N/Fe] and [Fe/H], as our result.
There are only two models that simultaneously reproduce both observed relations for our solar twins sample, but just roughly and qualitatively.
Like the others, these fiducial models entitled MWG-02 and MWG-07 are multi-zone models with two almost independent infall episodes,
a star formation rate proportional to both star and total surface mass density,
the Kroupa initial mass function (IMF) with a slope $x$\,=\,1.7 for the range of massive stars,
and a specific set of stellar yields for low-mass/intermediate-mass stars,
super-AGB stars (7-9\,M$_{\odot}$) and massive stars (see more details in \citet{Romano2019}).
The particular nucleosynthesis prescriptions of the MWG-02 model is based on
the stellar yields of low-mass/intermediate-mass stars from \citet{Karakas2010},
the stellar yields of massive stars from \citet{Nomotoetal2013} with no stellar rotation effects considered,
and absence of contributions by super-AGB stars, hypernovae and novae.
The ingredients of the MWG-07 model are:
stellar yields of low-mass/intermediate-mass stars, super-AGB stars and super solar metallicity stars from \citet{Venturaetal2013},
the stellar yields of massive stars from \citet{Limongi2018} with no stellar rotation effects included,
and absence of contributions by hypernovae and novae.

\citet{Nissen2015} also obtained a linear correlation of [C/Fe] as a function of age for a smaller sample of solar twins,
whose slope (0.0139$\pm$0.0020\,dex.Gyr$^{-1}$) and intercept (-0.110$\pm$0.011\,dex) are very close to our results.
Similarly to \citet{Nissen2015}, \citet{DaSilva2015} did an important work,
presenting for 140 FGK dwarfs a series of linear fits of abundance ratios [X/Fe]
as a function of [Fe/H] and isochrone stellar age.
Since \citet{DaSilva2015} performed independent fits for ages below and above the solar age without publishing the slopes,
we can just make a qualitative comparison.
Whilst they found a nearly global positive trend of [C/Fe] versus age, agreeing with our prediction,
their data suggest an overall negative trend for [N/Fe] as a function of age, opposite to our global single fit over the whole age scale, like
\citet{Suarez-Andres2016}, who also found negative trends of [N/Fe] as a function of age
for more restricted samples of solar-type stars with and without known planets in comparison with \citet{DaSilva2015}.
On the other hand, we have found a negative trend of [N/Fe] with age only for the solar twins older than the Sun.

Very recently, \citet{Nissenetal2020} have proposed that the relations between several abundance ratios and stellar age
derived from observations of solar type-stars ([C/Fe] and [Fe/H] included)
might be represented by two distinct sequences that would be interpreted as evidence of two episodes of star formation
induced by the accretion of gas onto the Galactic disc.
The separation of the two sequences in the stellar age scale would be around 5-6\,Gyr,
approaching close to the formation epoch of the Sun
(see Figure 3 in their paper showing the age-metallicity relation).
This result would substantiate the fits of [C,N/Fe] as a function of stellar age 
for FGK dwarfs younger and older than the Sun done by \citet{DaSilva2015}
and alternatively done in the current work too,
or even of [C,N/Fe] as a function of [Fe/H] by \citet{DaSilva2015},
and [C/Fe] as a function of [Fe/H] by \citet{Suarez-Andres2017}
for stars with [Fe/H] below and above zero.

We are able then to qualitatively compare our thin disc solar twins relation [C/Fe]-age
against the results from \citet{Nissenetal2020} obtained very recently
for 72 nearby solar-type stars with -0.3\,$\leq$\,[Fe/H] \,$\leq$\,+0.3\,dex through the analysis of HARPS spectra.
They split their sample into two distinct stellar populations based on their age-[Fe/H] distribution: old stars and young stars.
\citet{Nissenetal2020} found [C/Fe] increasing with the stellar age, like we have derived too.
A linear fit would not be good enough to reproduce the whole data distribution of both populations,
which cover ages from 0 up to 11\,Gyr.
None kind of fit was tried by them (rather than a linear fit,
perhaps, a higher order polynomial fitt would work for their data distribution).
Coincidently, they found that [C/Fe] tends to decrease from about +0.1\,dex 10\,Gyr ago
down to around -0.1\,dex now, such that the carbon-to-iron ratio
would be very close to the solar value nearly to 8.6\,Gyr ago, like we observe. 

Our relations [C/Fe]-[Fe/H] and [C/Fe]-age derived for local thin disc solar twins can also be compared
with the very recent carbon abundance measurements obtained for 2133 FGK dwarfs by \citet{Franchinietal2020} in the Gaia-ESO Survey.
They classified their sample stars as thin and thick disc members by using independent chemical, kinematical and orbital criteria.
The under-solar anti-correlation of  [C/Fe] versus [Fe/H] found by us qualitatively agrees
with the \citet{Franchinietal2020}'s relation obtained for thin disc dwarfs,
specially in the cases of the kinematical and orbital classifications
(unfortunately they did not apply any kind of fit to their data).
The overall increase of [C/Fe] as a function of age found by us
is compatible with the \citet{Franchinietal2020}'s results for the thin disc stars, which cover ages from 2 up to 12\,Gyr
(again, the kinematical and orbital classifications provide better agreements).
Differently from our analysis, the variation of the carbon-to-iron ratio
seems to better follow a high-order polynomial function of age rather than a linear fit.
They obtained [C/Fe] increasing from about -0.12\,dex 2\,Gyr ago up to near the solar level 12\,Gyr ago.

\subsection{$^{12}$C/$^{13}$C versus [C/Fe], [N/Fe], [Fe/H] and age}

\begin{figure}
\includegraphics[scale=0.300]{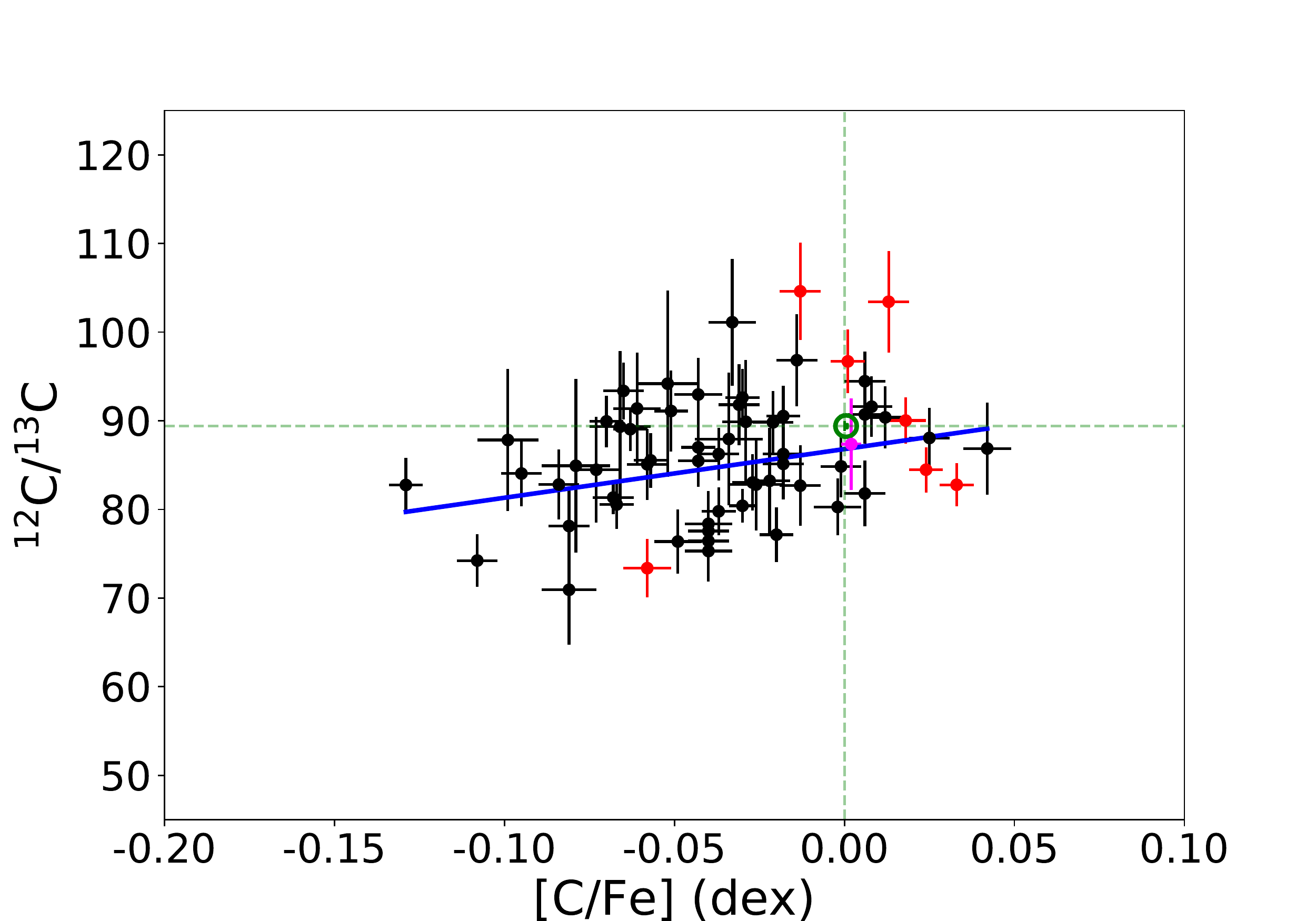}
\includegraphics[scale=0.300]{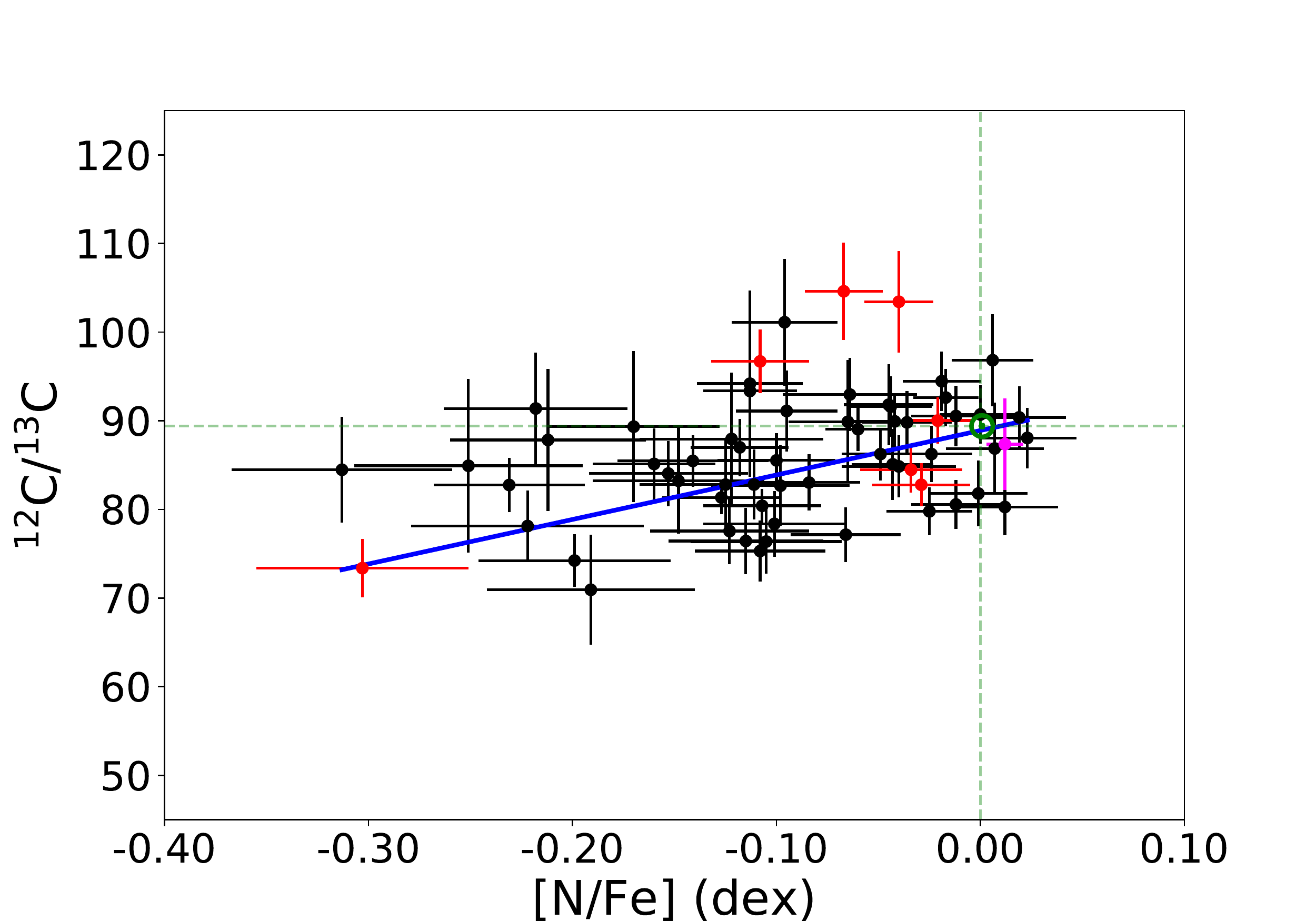}
\caption{
$^{12}$C/$^{13}$C as a function of [C/Fe] and [N/Fe] for the 55 thin disc solar twins.
The same description of Fig.~\ref{cfe-nfe-feh-age} is adopted.
The results of each linear fit are shown in Table~\ref{value-adj-razao}.
}
\label{razao-cfe-nfe}
\end{figure}

Since we have homogeneously measured the abundances of C, N and the isotopic ratio $^{12}$C/$^{13}$C,
we investigate effects due to some internal mixing process associated with the CNO cycles
that could have changed these photospheric chemical abundances,
perhaps altering their pristine chemical composition.
According to the plots of $^{12}$C/$^{13}$C versus [C/Fe] and [N/Fe] (Fig.~\ref{razao-cfe-nfe}),
the $^{12}$C/$^{13}$C ratio is somehow correlated with [C/Fe]
(slope with a relative error of 25\,per\,cent, i.e. 3.9$\sigma$ of significance),
as expected from some mixing process connected with a CNO cycle.
On the other hand, the C isotopic ratio is not anti-correlated with [N/Fe],
instead being positively correlated (slope with a relative error of about 16\,per\,cent),
unlike expectations for CNO processing.
These trends are compatible with no deep internal mixing in association with products from the CNO cycle.

\begin{table*}
\centering
\caption{Slopes, intercepts and $rms$ of the linear fits of $^{12}$C/$^{13}$C versus [C/Fe] and [N/Fe],
derived for the 55 thin disc solar twins.}
\label{value-adj-razao}
\begin{tabular}{llll|lll}
\hline
           & \multicolumn{3}{c}{{[}C/Fe{]}}              & \multicolumn{3}{c}{{[}N/Fe{]}}             \\
\hline
           & slope            & intercept       & $rms$   & slope            & intercept    & $rms$   \\
           & (dex$^{-1}$)     &                  &       & (dex$^{-1}$)     &                  &       \\
\hline
$^{12}$C/$^{13}$C  & 55$\pm$14  & 86.8$\pm$0.8 & 6.0 & 50.2$\pm$8.2  & 88.9$\pm$0.8 & 6.3 \\
\hline
\end{tabular}
\end{table*}

\begin{figure}
\includegraphics[scale=0.300]{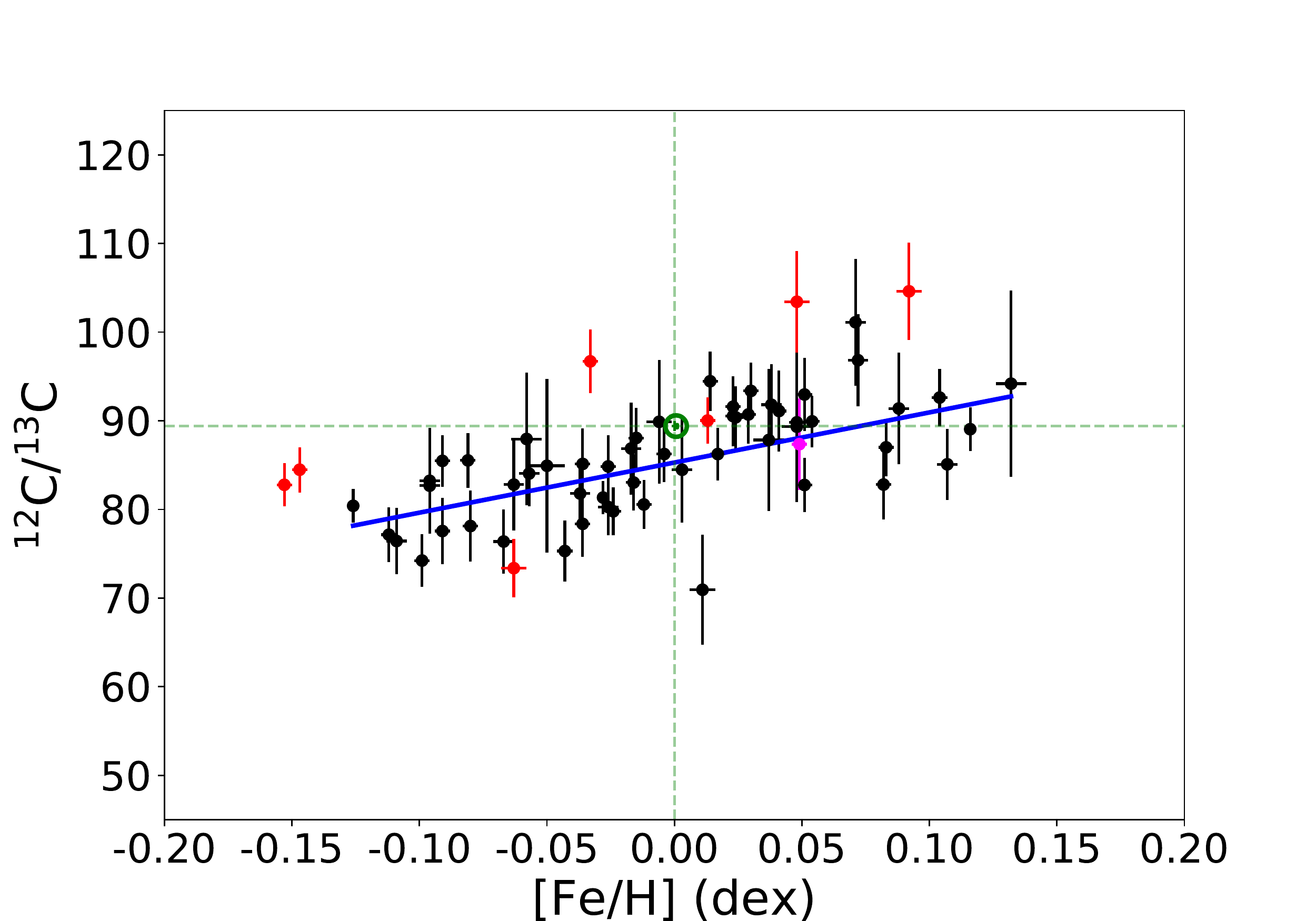}
\includegraphics[scale=0.300]{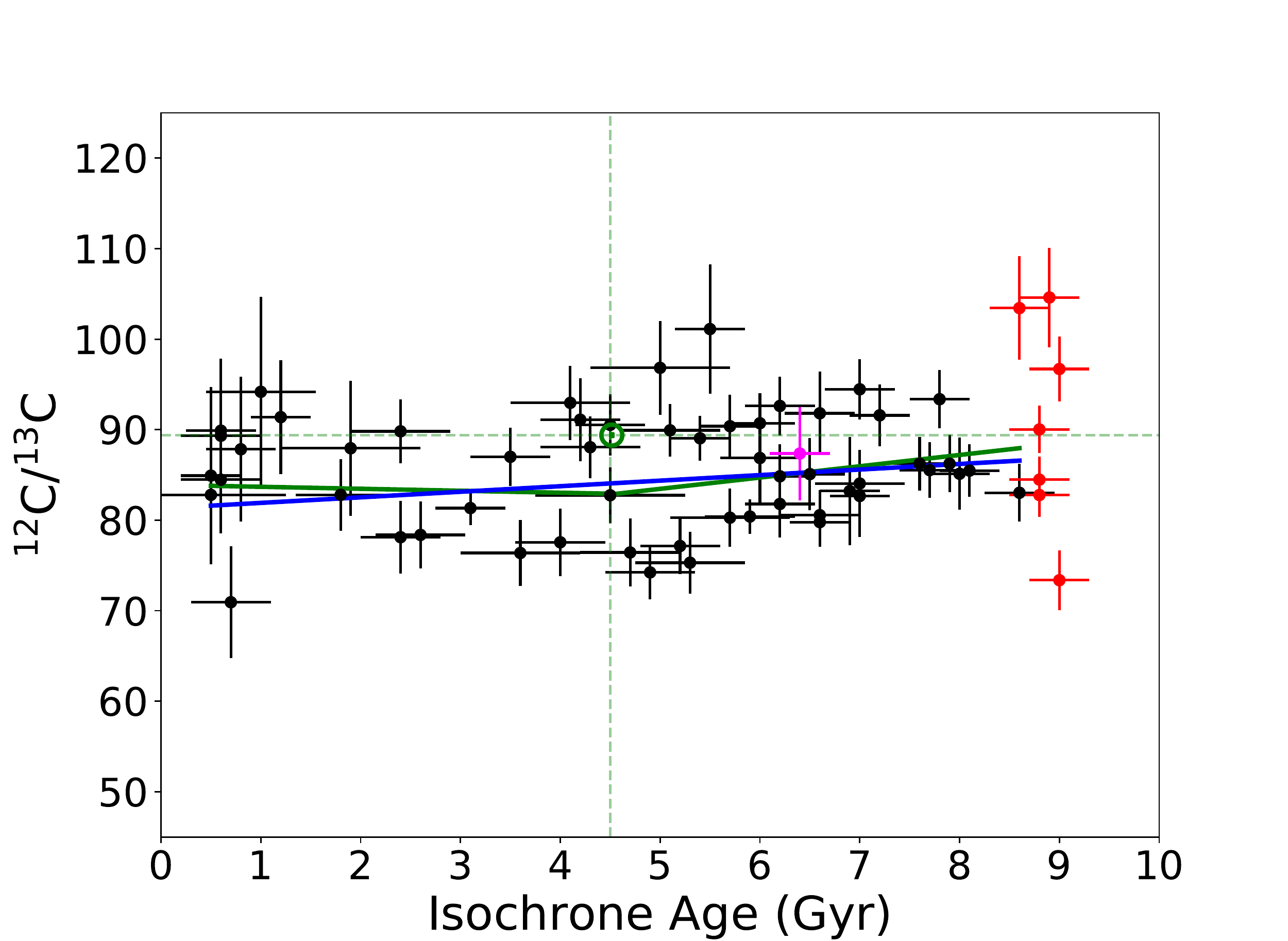}
\caption{
$^{12}$C/$^{13}$C as a function of [Fe/H] and age for the 55 thin disc solar twins.
The same description of Fig.~\ref{cfe-nfe-feh-age} is adopted.
The results of each linear fit are shown in Table~\ref{value-adj}.
}
\label{razao-feh-age}
\end{figure}

The linear fit of $^{12}$C/$^{13}$C as a function of metallicity (Fig.~\ref{razao-feh-age}) demonstrates that
$^{12}$C/$^{13}$C is positively well correlated with [Fe/H] for the solar twins sample
in their metallicity range -0.126\,$\leq$\,[Fe/H]\,$\leq$\,+0.132\,dex
(slope of +56.5$\pm$7.2\,Gyr$^{-1}$, i.e. slope with a relative error of just 13\,per\,cent).
The isotopic ratio $^{12}$C/$^{13}$C increases from 78.2 up to 92.8
in the interval of metallicity of our sample stars (a variation of 3\,$rms$ of the derived fit).
In fact, the statistical significance of this observational result
is basically on the overall increase of the isotopic ratio as a function of [Fe/H],
putting the absolute extreme values aside.
Surprisingly, the GCE models follow an opposite trend with [Fe/H],
something that is worth exploring in further GCE models.

The linear fit of $^{12}$C/$^{13}$C as a function of the isochrone stellar age (Fig.~\ref{razao-feh-age})
shows that $^{12}$C/$^{13}$C is marginally correlated with age for the solar twins sample
(slope of +0.614$\pm$0.250\,Gyr$^{-1}$, i.e. slope with a relative error of 41\,per\,cent, or just 2.5$\sigma$ of significance).
There seems to be a positive trend between this isotopic ratio and the isochrone age,
presenting overall decrease of $^{12}$C/$^{13}$C from a value of 86.6 (8.6\,Gyr ago)
down to 81.3 in the current time (a variation of only 0.8\,$rms$ of the derived fit).
The broken linear fit splitting the sample into solar twins younger and older than the Sun
is statistically equivalent to the global single fit (also see Tab.~\ref{value-adj}).

The result from the $^{12}$C/$^{13}$C-age relation is very interesting indeed in association with the other derived relation $^{12}$C/$^{13}$C-[Fe/H].
We remind the reader that the fundamental statistical significance of our result is focused on
the negative trend found between the isotopic ratio and the time
rather than the absolute values of the predictions for the local ISM now and 8.6\,Gyr ago.
The GCE models must prove or not this observational evolutive trend.
Nevertheless, the decrease of  $^{12}$C/$^{13}$C in time would be perharps expected along the evolution of the Galactic disc
due to the delayed contribution of the low-mass/intermediate-mass AGB stars.

The terrestrial $^{12}$C/$^{13}$C\,=\,89.4$\pm$0.2 \citep{Coplen2002},
as suggested by \citet{Asplund2009}
to represent the current solar value (and his pristine value too),
is in between the solar photosphere values of
86.8$\pm$3.8 by \citet{Scott2006},
91.4$\pm$1.3 by \citet{Ayres2013},
and 93.5$\pm$0.7 by \citet{Lyons2018}.
Possible causes for the small apparent discrepancy between the terrestrial and solar ratios are discussed by \citet{Lyons2018}.

Recently, the $^{12}$C/$^{13}$C ratio has been employed by \citet{Adibekyan2018}
as one of the criteria defining solar siblings, meaning stars that formed in the same birth cluster as the Sun,
being thus stars with ages, chemical compositions and kinematics as the Sun \citep{Ramirez2014b}.
We recommend that the variation in the $^{12}$C/$^{13}$C ratio around the solar value
should be within 4.5 unities: 84.9\,$\leq$\,$^{12}$C/$^{13}$C\,$\leq$\,93.9
(i.e. with a 3$\sigma$ agreement around the solar value proposed by \citet{Asplund2009} and \citet{Coplen2002},
assuming an average error from the four values previously listed).

We have found 28 stars in our sample of 55 thin disc solar twins
that could be solar sibling candidates based on the C isotopic ratio only.
However, considering that solar siblings must have similar ages to the Sun, only 10 stars are left
(HIP\,025670, HIP\,040133, HIP\,049756, HIP\,064673, HIP\,07585, HIP\,079672, HIP\,089650, HIP\,095962, HIP\,104045 and HIP\,117367).
We propose that the age interval for the convergence to the solar age should vary from 3.36\,Gyr up to 5.76\,Gyr,
i.e. 3$\sigma$ around the solar value assuming the typical error in the isochrone age of our solar twins sample. 
Furthermore, nevertheless, the stars must have similar kinematics to the Sun,
and further applying this condition, none solar sibling candidate remains as solar sibling.

Concerning the comparisons against predictions of the state-of-the-art GCE models
to track the CNO isotopes in the local ISM by \citet{Romano2017},
our results for solar twins, spanning ages between 0.5 up to 8.6\,Gyr (around 8\,Gyr in time),
agree with the predictions of one of their models,
because the $^{12}$C/$^{13}$C ratio appears actually decreasing since 8.6\,Gyr up to now (i.e. since 5\,Gyr ago until nowadays)
and this model simultaneously reproduce both the solar and ISM $^{12}$C/$^{13}$C ratios.
This model does not include C yields from super-AGB stars
and does not take the effects of stellar rotation on the yields of massive stars into account 
(Model\,1 in \citet{Romano2017} renamed as MWG-02 in \citet{Romano2019}).
Similarly to the \citet{Romano2019}'s GCE models,
it is a multi-zone model with two infall episodes,
star formation rate proportional to both star and total surface mass density,
the Kroupa IMF under a slope $x$\,=\,1.7 for the range of massive stars, and
a set of stellar yields for low-mass/intermediate-mass stars, super-AGB stars and massive stars (read details in \citet{Romano2017}).

Whilst the current C isotopic ratio for the ISM of the neighbourhood adopted by \citet{Romano2017} is 68$\pm$15
(as measured by \citet{Milam2005} as a typical for the local ratio),
we suggest a value of 81.3 ($\pm$1.4),
such that there is an agreement within 1$\sigma$.

\subsection{[C/N], [C/O] and [N/O] versus [Fe/H], [O/H] and age}

\begin{figure*}
\includegraphics[scale=0.315]{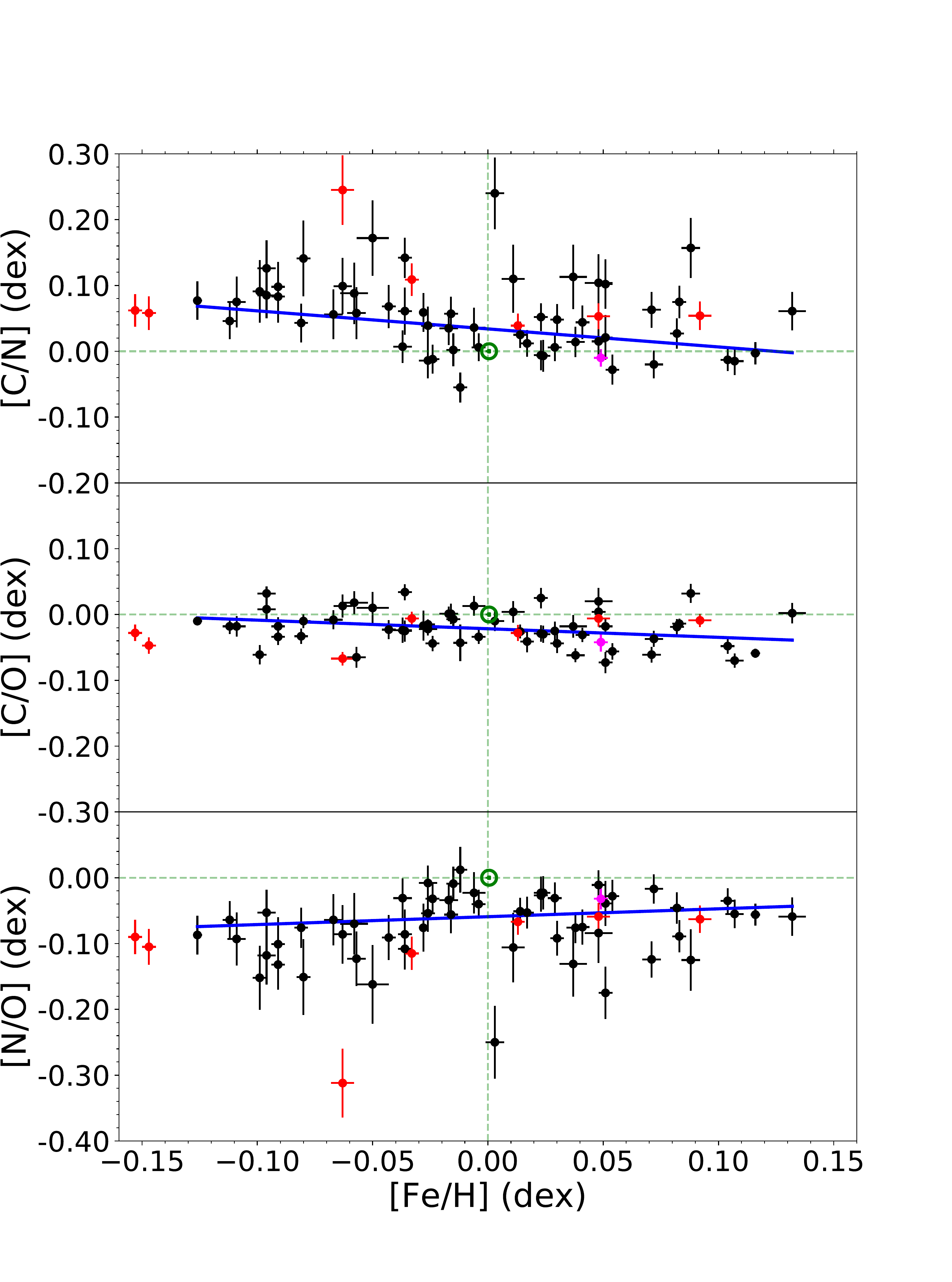}
\includegraphics[scale=0.315]{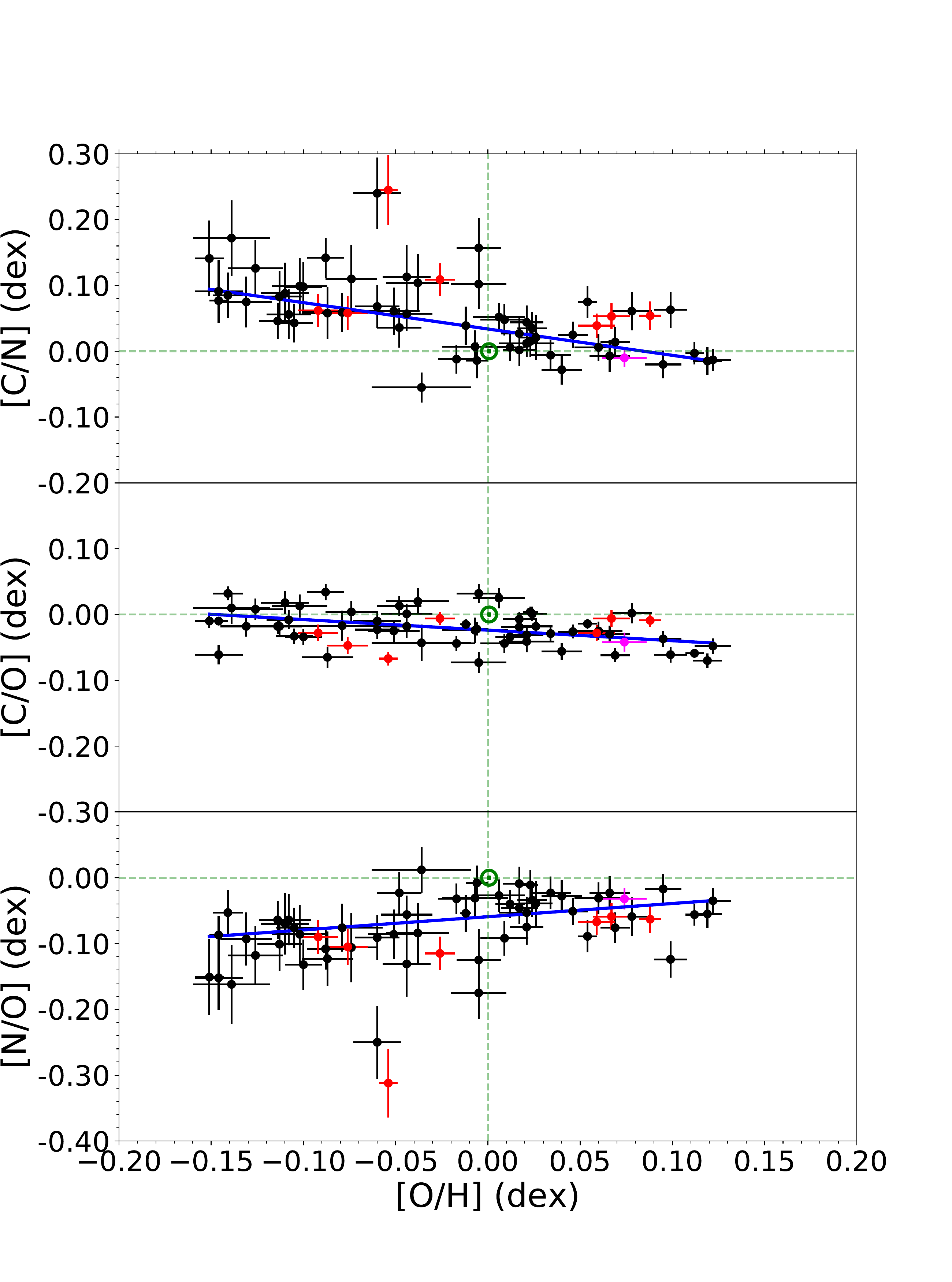}
\caption{
[C/N], [C/O] and [N/O] as a function of [Fe/H] and [O/H] for the 55 thin disc solar twins.
The same description of Fig.~\ref{cfe-nfe-feh-age} is adopted.
The results of each linear fit are shown in Table~\ref{value-adj} and Table~\ref{value-adj-oh}.
}
\label{artigo_cn_co_no_feh_oh}
\end{figure*}

We have used our measurements of carbon and nitrogen
together with the oxygen abundances homogeneously derived by \citet{Bedell2018} for the same solar twins sample,
for computing their [C/N], [C/O] and [N/O] ratios.
Regarding the relations of these abundance ratios as a linear function of [Fe/H]
(shown in Fig.~\ref{artigo_cn_co_no_feh_oh}),
we have found that:
{\bf (i)} [C/N] is anti-correlated with [Fe/H]
(the relative error of the negative slope is 22\,per\,cent, i.e. 4.6$\sigma$ of significance level);
{\bf (ii)} [C/O] is anti-correlated with [Fe/H]
(relative error of the negative slope of about 18\,per\,cent); and
{\bf (iii)} there is a tentative correlation between [N/O] and [Fe/H]
(roughly 2$\sigma$ significance level).
The Sun, when compared against the fits versus [Fe/H] of the solar twins,
seems to be about normal (within about 1-sigma) in the C/N, C/O and N/O ratios.
In \citet{Melendez2009}, the [C,N,O/Fe] ratios are about the same within the error bars,
which is expected considering that the three elements have low condensation temperatures.
In this sense, the C/N, C/O and N/O ratios in \citet{Melendez2009} 
are roughly about the same (within the errors) in the Sun and solar twins,
as also found in the present work.

Due to an existence of a few electronic transitions of oxygen in the optical range
and many Fe\,I and Fe\,II lines in the case of  FGK stars,
iron is usually used to trace the stellar metallicity.
On the other hand, oxygen is the most abundant metal in stars,
being in fact considered the best indicator of the metallicity of any star or stellar system.
Therefore, it becomes interesting to compare [C/N], [C/O] and [N/O] directly with [O/H],
also because these abundance ratios measured in the gas phase for other galaxies are often given as a function of O
and, secondly, oxygen has a simpler nucleosynthetic origin
(basically made by massive stars that die as SN-II) in comparison with iron (both SN-II and SN-Ia).

[C/N], [C/O] and [N/O] as a function of [O/H] for the 55 thin disc solar twins are also shown in Fig.~\ref{artigo_cn_co_no_feh_oh}.
Regarding the correspondent linear fits, we have found that:
{\bf (i)} [C/N] is anti-correlated with [O/H] 
(the negative slope has 7.9$\sigma$ of significance level);
{\bf (ii)} [C/O] is anti-correlated with [O/H] (negative slope with 8.1$\sigma$ significance level);
and {\bf (iii)} [N/O] is correlated with [O/H] (positive slope with 3.9$\sigma$ significance level).

[C/N]  and [C/O] present the same behaviour as a function of both [Fe/H] and [O/H],
but while [N/O] is well correlated with [O/H], it only shows a tentative correlation against [Fe/H].
The Sun seems normal (within about 1-sigma) in the C/N, C/O and N/O ratios.

Making an analogy between the $\alpha$-elements oxygen and magnesium  (products of the evolution of massive stars),
we can make a qualitative comparison between our relation [C/O]-[O/H] derived for local thin disc solar twins 
and the global relation [C/Mg]-[Mg/H] observed by the GALAH\footnote{GALactic Archaeology with HERMES}
optical spectroscopic survey for two distinct Galactic's disc stellar populations
that embraces 12,381 out of 70,924 GALAH stars \citep{Griffithetal2019}.
Their two stellar groups are denominated by high-$\alpha$/low-Ia and low-$\alpha$/high-Ia (where Ia means SN-Ia)
based on cuts in [Mg/Fe] in the diagram [Mg/Fe] vs. [Fe/H] (see their Figure 2),
corresponding, respectively, to thick and thin disc stars as chemically identified by others works (e.g. \citet{Franchinietal2020}).
\citet{Griffithetal2019} found that [C/Mg] is anti-correlated with [Mg/H] for low-$\alpha$ (thin disc members), although no fit is provided.
[C/Mg] versus [Mg/H] qualitatively shows the same behaviour of our fit [C/O]-[O/H], albeit their relation seems steeper.

By  {\lq\lq}cross-correlating{\rq\rq} our observed relations [C/O] and [N/O] versus [O/H] with the predictions of the GCE models by \citet{Romano2019},
we have found there is no agreement in the case of the [C/O]-[O/H] relation,
observed by us as an anti-correlation, differently from a positive correlation predicted by some \citet{Romano2019}'s models
covering the chemical composition of solar twins ([C/O]\,=\,[O/H]\,=\,0.0($\pm$\,0.15)\,dex).
On the other hand, there is a qualitative agreement of the observed relation [N/O]-[O/H] for our solar twins sample
against some \citet{Romano2019}'s models, showing that [N/O] is positively correlated with [O/H] around [N/O]\,=\,[O/H]\,=\,0.0($\pm$\,0.13)\,dex.
The \citet{Romano2019}'s MWG-07 model,
which roughly and simultaneously reproduces both observed relations [C,N/Fe]-[Fe/H] (Subsection\,4.1),
is not among these models that show qualitative agreements against the [N/O]-[O/H] observed relation.
However, one of them is coincidently the same model that reproduces our observed relation $^{12}$C/$^{13}$C-age,
which is named as MWG-02 by \citet{Romano2019} and is equivalent to the Model\,1 in \citet{Romano2017}.

\begin{figure}
\includegraphics[scale=0.400]{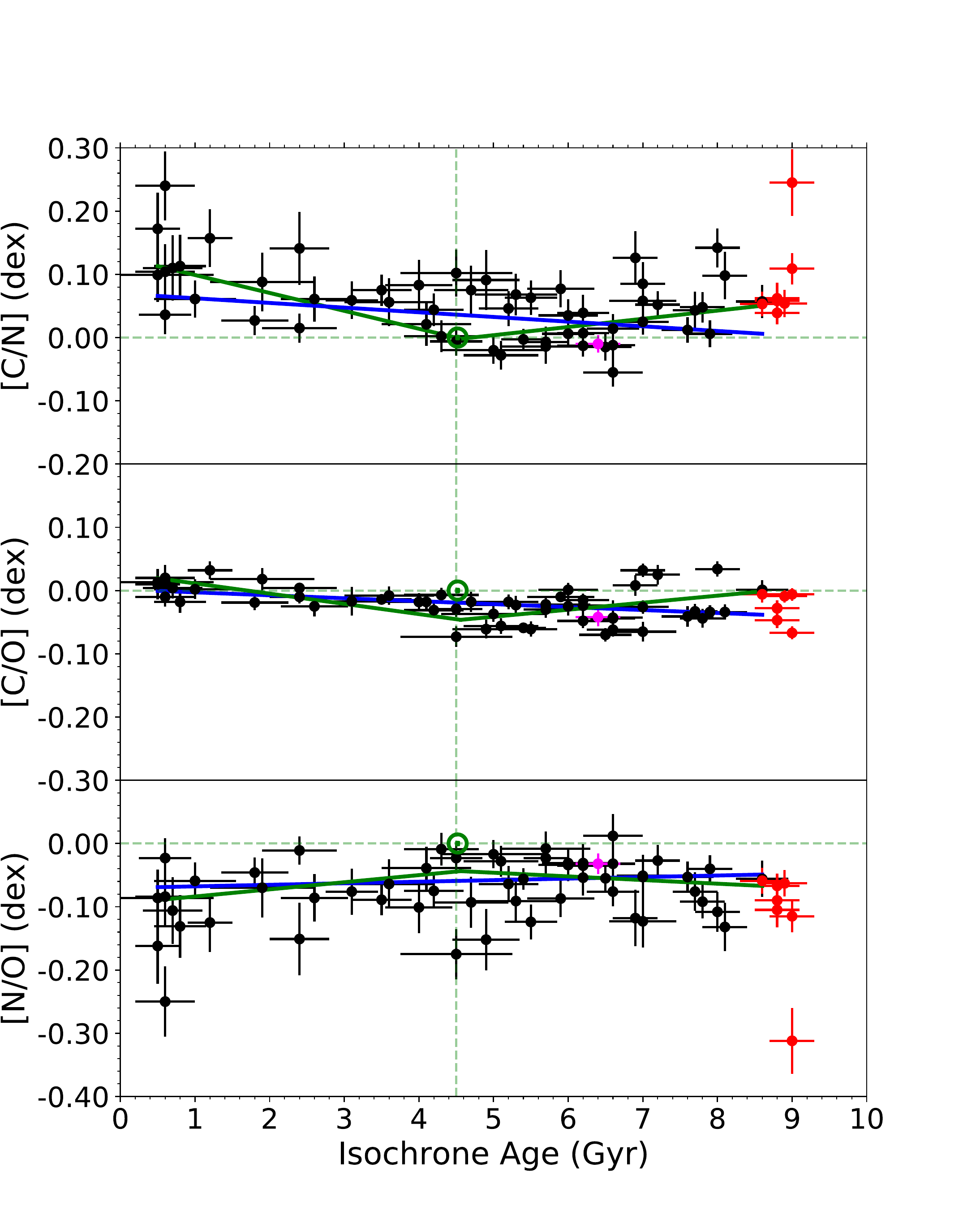}
\caption{
[C/N], [C/O] and [N/O] as a function of the isochrone age for the 55 thin disc solar twins.
The same description of Fig.~\ref{cfe-nfe-feh-age} is adopted.
The results of each linear fit are shown in Table~\ref{value-adj}.
}
\label{artigo_cn_co_no_age}
\end{figure}

\begin{table*}
\centering
\caption{
Slopes, intercepts and $rms$ of the global linear fits of
[C/Fe], [N/Fe], [C/N], [C/O], [N/O] and $^{12}$C/$^{13}$C
versus [Fe/H] and isochrone stellar age,
derived for the 55 thin disc solar twins (first row of each abundance ratio).
These parameters are also listed for the broken linear fits carried out only as a function of age
that split the sample into solar twins younger and older than the Sun (respectively second and third rows of each abundance ratio).
}
\label{value-adj}
\begin{tabular}{lrrrrrr}
\hline
           & \multicolumn{3}{c}{{[}Fe/H{]}}              & \multicolumn{3}{c}{stellar age}             \\
\hline
           & slope            & intercept     & $rms$ & slope            & intercept     & $rms$   \\ 
           &                       & (dex)            & (dex) & (dex.Gyr$^{-1}$) & (dex)            & (dex) \\
\hline
{[}C/Fe{]} & -0.056$\pm$0.012 & -0.039$\pm$0.001 & 0.033 &  0.013$\pm$0.001 & -0.107$\pm$0.003 & 0.037 \\
                 &                                   &                                   &             &  0.007$\pm$0.004 & -0.093$\pm$0.014 & 0.038 \\
                 &                                   &                                   &             &  0.022$\pm$0.006 & -0.160$\pm$0.036 & 0.040 \\
{[}N/Fe{]} &  0.162$\pm$0.057 & -0.064$\pm$0.003 & 0.080 &  0.013$\pm$0.002 & -0.135$\pm$0.010 & 0.070 \\
                 &                                   &                                   &             &  0.041$\pm$0.008 & -0.207$\pm$0.031 & 0.075 \\
                 &                                   &                                   &             & -0.015$\pm$0.007 &  0.045$\pm$0.059 & 0.059 \\
{[}C/N{]}  & -0.274$\pm$0.059 & 0.034$\pm$0.004  & 0.056  & -0.007$\pm$0.002 &  0.069$\pm$0.011 & 0.054 \\
                 &                                   &                                   &             & -0.028$\pm$0.006 &  0.127$\pm$0.023 & 0.052 \\
                 &                                   &                                   &             &  0.013$\pm$0.006 & -0.061$\pm$0.045 & 0.046 \\
{[}C/O{]}  & -0.130$\pm$0.023 & -0.022$\pm$0.002 &  0.026 & -0.005$\pm$0.001 &  0.002$\pm$0.004 & 0.025 \\
                 &                                   &                                   &             & -0.016$\pm$0.003 &  0.027$\pm$0.012 & 0.018 \\
                 &                                   &                                   &             &  0.011$\pm$0.004 & -0.097$\pm$0.025 & 0.027 \\
{[}N/O{]}  &  0.120$\pm$0.061 & -0.059$\pm$0.004 & 0.050  &  0.002$\pm$0.002 & -0.070$\pm$0.011 & 0.050 \\
                &                                    &                                   &             &  0.011$\pm$0.005 & -0.095$\pm$0.020 & 0.057 \\
                &                                    &                                   &             & -0.006$\pm$0.005 & -0.017$\pm$0.039 & 0.040 \\
\hline
                                    & (dex$^{-1}$)                          &                  &       & (Gyr$^{-1}$)     &                  &       \\
\hline
$^{12}$C/$^{13}$C & 56.5$\pm$7.2 & 85.3$\pm$0.5  & 4.7 &  0.614$\pm$0.250 & 81.3$\pm$1.4     & 6.3 \\
                                    &                           &                            &        & -0.218$\pm$0.893 & 83.9$\pm$3.4 & 6.2 \\
                                    &                           &                            &        & 1.248$\pm$0.709  & 77.2$\pm$6.2 & 6.3 \\
\hline
\end{tabular}
\end{table*}

\begin{table}
\centering
\caption{
Slopes, intercepts and $rms$ of the global linear fits of [C/N], [C/O], [N/O] versus [O/H],
derived for the 55 thin disc solar twins.
}
\label{value-adj-oh}
\begin{tabular}{lrrrrrr}
\hline
           & \multicolumn{3}{c}{{[}O/H{]}}        \\
\hline
           & slope    & intercept & $rms$   \\ 
           &               & (dex)             & (dex)  \\
\hline
{[}C/N{]}  & -0.400$\pm$0.051 &  0.034$\pm$0.004 & 0.046 \\
{[}C/O{]}  & -0.161$\pm$0.020 & -0.024$\pm$0.002 & 0.025 \\
{[}N/O{]}  &  0.201$\pm$0.053 & -0.059$\pm$0.004 & 0.045 \\
\hline
\end{tabular}
\end{table}

After inspecting the global linear relations of [C/N], [C/O] and [N/O] as a function of the isochrone stellar age
in the wide range of ages of the 55 thin disc solar twins (Fig.~\ref{artigo_cn_co_no_age}),
we have found that:
{\bf (i)} [C/N] is somehow anti-correlated with age
(the relative error of the negative slope is about 29\,per\,cent, i.e. 3.5$\sigma$ of significance level);
{\bf (ii)} [C/O] is anti-correlated with age (a relative error of the negative slope of about 20\,per\,cent);
and {\bf (iii)} there is no correlation between [N/O] and age
(the relative error of the slope reaches 100\,per\,cent).

Alternatively, we have also derived composed linear fits of [C/N], [C/O] and [N/O] by splitting the sample into solar twins younger and older than the Sun
(see the parameters of the broken linear fits in Tab.~\ref{value-adj} and the plots in Fig.~\ref{artigo_cn_co_no_age}).
The [C/N] and [C/O] ratios are found to be anti-correlated with the stellar age for the solar twins younger than the Sun
(slightly more anti-correlated than in the global single fits),
but both ratios show a positive trend with age for the older solar twins.
Curiously, the break point of the [C/N]-age broken fit matches the solar ratio.
[N/O], as shown by the broken fit for {\lq\lq}young{\rq\rq} and {\lq\lq}old{\rq\rq} solar twins,
follows the same overall behaviour of the global single linear fit (i.e. it is likely constant in time, in fact).
The overall dispersion of every broken fit is comparable to the dispersion of the correspondent global single fit.
Notice that dependence of the [C/N] and [C/O] ratios on age (or temporal evolution of the thin disc) is enhanced for the {\lq\lq}young{\rq\rq} solar twins
in comparison with the overall behaviour derived from the single linear fit (the slopes are different in 2-sigma).
The result for [C/O] corroborates the rise of [C/Mg] for thin disc young stars as observed by \citet{Franchinietal2020},
suggesting two competing scenarios:
{\bf (i)} a delayed contribution from low-mass stars to explain the carbon abundance
in the Galactic thin disc along the recent times \citep{Marigoetal2020},
and {\bf (ii)} an enhanced mass loss through stellar winds from metal-rich massive stars
that produces an increased C yield at solar and super-solar metallicities.

On one hand, the Sun, when compared against the fits versus age of the solar twins,
seems about normal (within about 1-sigma) in the C/N, C/O and N/O ratios.
On the other hand, the C/N ratio in the Sun is placed close to the bottom end of the solar twins distribution
in all [C/N]-[Fe/H], [C/N]-[O/H] and [C/N]-age planes.
The solar C/O ratio is specifically found near to the distribution high end for the mid-age solar twins only
(all [C/O]-[Fe/H], [C/O]-[O/H] and [C/O]-age distributions),
that qualitatively agrees well with \citet{Bedell2018}, who analysed the same solar twins sample,
as well as with \citet{Nissenetal2020} for a solar analogs sample.
Particularly, the N/O ratio in the Sun is found to be placed at the high end of the overall solar twins distribution
(1.2 $rms$ in average above the relations [N/O]-[Fe/H], [N/O]-[O/H] and [N/O]-age of the solar twins).

Our derived relation [C/O]-age for thin disc solar twins can be finally and directly compared
with the relations [C/Mg] vs. age and C/O number ratio vs. age by \citet{NissenGustafsson2018},
which compiled data from a set of different samples of solar twins:
\citet{Nissen2015}, \citet{Nissenetal2017} and \citet{Bedell2018}
(the former work handled the same sample of the current work, but performed their own C measurements as explained before).
While the \citet{NissenGustafsson2018}'s [C/Mg]-age relation has a large data dispersion making hard to assess a temporal analysis,
the C/O ratio appears to decrease toward solar twins younger than the Sun, differently from what we have observed.
However, the variation of the C/O ratio at a fixed age is comparable with the abundance dispersion measured by us.

Taking into account the global behaviour of these abundance ratios over the whole age scale, 
in fact the variations of [C/N] and [C/O] are both really small
along the evolution of the Galactic thin disc in the solar neighbourhood (only 10-17\,per\,cent),
and there is no variation at all for [N/O].
[C/N] in the ISM would have increased from around the solar value 8.6 Gyr ago up to +0.07\,dex now,
whilst [C/O] would have increased from -0.04 dex 8.6 Gyr ago up to the solar ratio now.

The absence of evolution of the N/O ratio in both time and [Fe/H] in solar twins
suggests uniformity in the nitrogen-oxygen budget for potential giant planet formation around these stars
(note that C/O has or would have small variations versus age and [Fe/H] respectively,
as also derived by \citet{Bedell2018} and \citet{Nissen2015}).

We are now able to state that
along the evolution of the Galactic local disc,
C seems to have been stochastically more accumulated in the ISM relatively compared with N and O,
since the abundance ratios C/N and C/O increase with time,
although they decrease with both metallicity indicators (Fe and O abundances) in the limited metallicity coverage of solar twins.
As expected, C and N are produced and ejected by low-mass/intermediate-mass (AGB phase) plus massive stars (SN-II),
and O basically by massive stars only.
Even though the N/O ratio is kept constant along the time,
N does not necessarily follow O specially because [N/O] is found to be correlated with [O/H].
Therefore, the N production seems
to have a more relative contribution from massive stars than carbon,
or, in other words, carbon would be relatively more synthesized by low-mass/intermediate-mass stars than nitrogen
(as suggested by \citet{Franchinietal2020} and \citet{Marigoetal2020}).
However, alternatively, an enhanced mass loss from metal-rich massive stars
could produce an increased C yield at solar and super-solar metallicities.

\section{Summary and Conclusions}

We have measured C, $^{12}$C/$^{13}$C and N in 63 solar twins through a self-consistent and homogenous procedure
based on a high-precision spectral synthesis of molecular lines/features in the blue region (specifically $\lambda\lambda$4170-4400\,{\AA}),
being 7 of them old alpha-enhanced solar twins and 1 having an excess of $s$-process elements,
which were excluded from further analysis,
remaining 55 solar twins for studying correlations with age and metallicity in the solar neighbourhood.
All 55 stars have thin disc kinematics \citep{Ramirezetal2012}.
The carbon abundance has been derived from $^{12}$CH\,A-X lines
and the $^{12}$C/$^{13}$C ratio from $^{13}$CH-$^{12}$CH features of the same electronic system.
The nitrogen abundance has been extracted from $^{12}$C$^{14}$N B-X lines (CN Violet System),
taking into account the measured carbon abundance and its main isotopic ratio.
We provide comprehensive lists of these molecular lines and features: twelve CH\,A-X lines, six $^{13}$CH-$^{12}$CH\,A-X features and five CN\,B-X lines.
The average errors in the C and N abundances are around 0.006 and 0.03\,dex respectively (very similar in both [X/Fe] and [X/H] scales).
Specifically, the error in $^{12}$C/$^{13}$C changes from 1.9 up to 10.5 (about 4.3 as a mean value).

We could obviously conclude, as expected, that the measured abundances of C and N as well as the isotopic ratio $^{12}$C/$^{13}$C
represent in fact the pristine chemical composition of every solar twin of our sample,
because we have not found an overall connection among the photospheric $^{12}$C/$^{13}$C, [C/Fe], [N/Fe] and [C/N]
that could be an indicative of some internal mixing process associated with the CNO cycles
(for instance, $^{12}$C/$^{13}$C is not correlated with [$^{12}$C/$^{14}$N], being anti-correlated indeed).

We confirm that the Sun is slightly enhanced in the volatiles C and N relatively to Fe in comparison with solar twins,
like \citet{Bedell2018}, \citet{Nissen2015}, and \citet{Melendez2009} have already found for volatiles in general.
On the one hand, the Sun seems to be normal in the $^{12}$C/$^{13}$C, C/O, C/N and N/O ratios within the errors (about 1-sigma) in comparison with solar twins.
Notice that \citet{Melendez2009} found comparable (within the errors) C/Fe, N/Fe and O/Fe ratios in the Sun
relative to the solar twins, probably because the highly volatile elements C, N and O have similar low condensation temperatures among them.
On the other hand, we have particularly found that the N/O ratio in the Sun is placed at the high end of the overall solar twins distribution,
the solar C/N ratio just close to the distribution bottom,
and the solar C/O ratio near to the high end for the distribution of the mid-age solar twins only.

The linear fits of [C/Fe] and [N/Fe], as a function of [Fe/H] and isochrone stellar age for the analysed solar twins sample,
lay under solar for the whole scale of both [Fe/H] and age.
Whilst [C/Fe] decreases with [Fe/H] in the restrict metallicity interval of solar twins, we have found a positive trend between [N/Fe] and [Fe/H].
On the other hand, both [C/Fe] and [N/Fe] increase with age
(or, in other words, both [C/Fe] and [N/Fe] decrease with time from around the solar ratio value 8.6 Gyr ago down to nearly -0.1\,dex now).
Furthermore, the Sun is placed around 1$\sigma$ in average above the linear fits [X/Fe]-[Fe/H] and [X/Fe]-age.
This indicates that the Sun is slightly enhanced in both C and N relative to Fe in comparison with solar twins.
For the relations [C/Fe]-[Fe/H] and [N/Fe]-[Fe/H],
our results agree with \citet{DaSilva2015, Suarez-Andres2016, Suarez-Andres2017} that investigated solar-type dwarfs.
For the [C/Fe]-[Fe/H] relation, our results also agree with \citet{Nissen2015},
who analysed solar twins like the ones studied in this work.
Our under-solar anti-correlation of  [C/Fe] vs. [Fe/H] qualitatively agrees
with the \citet{Franchinietal2020}'s relation obtained for thin disc dwarfs
(specially in the cases of their kinematical and orbital classifications).
Regarding the [C/Fe]-age relation, our results agree with \citet{DaSilva2015, Bedell2018} (only this work done for solar twins).
Coincidently, \citet{Nissenetal2020} found that [C/Fe] tends to decrease from about +0.1\,dex 10\,Gyr ago
down to around -0.1\,dex now, such that the carbon-to-iron ratio
would be very close to the solar value nearly to 8.6\,Gyr ago,
like we observe through both kinds of linear fits (single and broken). 
The overall increase of [C/Fe] as a function of age found by us
is compatible with the \citet{Franchinietal2020}'s results that cover ages from 2 up to 12\,Gyr.
However, for the [N/Fe]-age relation, our results do not agree with neither \citet{DaSilva2015} nor \citet{Suarez-Andres2016},
perhaps because these studies were not restricted to solar twins.

Our solar twins sample has the advantage, relative to solar type dwarfs,
that a better precision can be obtained in the solar twins,
because systematic errors largely are canceled out in a differential analysis,
resulting in improved stellar parameters and chemical abundances,
and also in more precise stellar ages relative to the Sun, which is at near the middle of the age distribution of solar twins.

Comparisons against predictions of inhomogeneous chemical evolution models for the solar neighbourhood by \citet{Sahijpal2013}
show that our results agree, qualitatively, for the relations [C/Fe]-[Fe/H] and [N/Fe]-[Fe/H] around [Fe/H]=0 for all models,
because they predict [C/Fe]-[Fe/H] with negative slope and [N/Fe]-[Fe/H] with positive slope around the solar metallicity.
These results suggest that possibly C and N have different stochastic nucleosynthetic productions
during the evolution of the Galaxy's thin disc.

We have measured the $^{12}$C/$^{13}$C isotopic ratio for solar twins for the first time,
also focusing on the evolutive analysis on those 55 thin disc solar twins.
Our measurements are certainly useful as important constraints for chemical evolution models of the solar neighbourhood,
as \citet{Romano2017} had pointed out.
We predict 81.3 ($\pm$1.4) for the current C isotopic ratio in the ISM of the solar neighbourhood,
which is in agreement within 1$\sigma$ with the observed value of 68$\pm$15 by \citet{Milam2005}.

No solar sibling is found in our sample following the criteria
of similar ages, chemical compositions ($^{12}$C/$^{13}$C included) and kinematics
to the Sun \citep{Ramirez2014b, Adibekyan2018}.

We have found that $^{12}$C/$^{13}$C is positively well correlated with [Fe/H] in the small metallicity range of solar twins.
This result should be tested against predictions of robust GCE models for the solar vicinity.
Regarding specifically the $^{12}$C/$^{13}$C-age relation,
this isotopic ratio seems to have decreased a little bit in time along the evolution of the nearby Galaxy's thin disc.
A possible positive trend between $^{12}$C/$^{13}$C and stellar age has been found by us.
This is actually in qualitative agreement with the predictions of a couple of GCE models
 \citep{Romano2017, Romano2019}.

We have obtained linear anti-correlations for [C/N] and [C/O] as a function of [Fe/H] and [O/H].
For specifically [N/O], we have found no correlation with [Fe/H], but a positive correlation with [O/H].
We have derived linear anti-correlations for [C/N] and [C/O] as a function of age
(i.e. both abundance ratios increase with time), and again there is no correlation for [N/O].
[C/O] seems to be more anti-correlated with [Fe/H] and age than [C/N].
In fact, the variations of [C/N] and [C/O] are both really small
along the evolution of the Galactic thin disc in the solar neighbourhood.
However, we have found greater increase rates in time of both [C/N] and [C/O] ratios
for the solar twins younger than the Sun in comparison with the overall single fit.

We have found one common GCE model by \citet{Romano2017} and \citet{Romano2019} (under different denominations),
which roughly, qualitatively and simultaneously predict
the relations [C/Fe]-[Fe/H], [N/Fe]-[Fe/H], $^{12}$C/$^{13}$C-age and [N/O]-[O/H]
derived in the present work for a sample of thin disc solar twins.
Surprisingly, the GCE models follow an opposite trend between $^{12}$C/$^{13}$C and [Fe/H] (we have obtained a direct correlation),
something that is worth exploring in further GCE models.
Moreover, GCE models must also prove or not the observational trend between $^{12}$C/$^{13}$C and stellar age derived by us for solar twins.

We can conclude that C does not exactly follow neither N nor O as a function of [Fe/H] and [O/H] and the time too.
We can also state that carbon and nitrogen likely have had different nucleosynthetic origins along the Galaxy thin's disc evolution.
The N production seems to have a more relative contribution from massive stars than carbon,
or, in other words, carbon would be relatively more synthesized by low-mass/intermediate-mass stars than nitrogen
(as suggested by \citet{Franchinietal2020} and \citet{Marigoetal2020}).
Particularly, we have found an increase in time of both [C/N] and [C/O] ratios
for the solar twins younger than the Sun.
However, an alternative explanation could come from an enhanced mass
loss through stellar winds from metal-rich massive stars
that would increase the C yield at solar and super-solar metallicities.

The absence of evolution of [N/O] in both [Fe/H] and time in solar twins
suggests uniformity in the nitrogen-oxygen budget for potential giant planet formation around these stars,
although [N/O] is seen to be correlated with [O/H].
Our results can contribute to the {\lq}CNO composition -- planet formation{\rq} connection
(mainly linked to the formation of icy planetesimals, watery super-earths and/or giant planets, \citet{Marboeufetal2014}).

Our results for C, $^{12}$C/$^{13}$C and N in a sample of nearby solar twins,
with well-determined parameters and spanning a wide range in age,
are certainly excellent constraints to understand the chemical evolution of thin disc in the solar vicinity,
as well as to study the nucleosynthetic
origins of C, N, O and Fe along the Galaxy's evolution.

\section*{Acknowledgements}

This study was financed in part
by the Coordena\c c\~ao de Aperfei\c coamento de Pessoal de N\'\i vel Superior - Brasil (CAPES) - Finance Code 001
(RBB and ADCM acknowledge this support).
ADCM also thanks Conselho Nacional de Desenvolvimento Cient\'\i fico e Tecnol\'ogico (CNPq)
for the research productivity grant (309562/2015-5).
JM thanks FAPESP (2018/04055-8) and CNPq for the research productivity grant (306730/2019-7).
LS acknowledges financial support from the Australian Research Council (Discovery Project 170100521).
We are also grateful for the anonymous referee by his/her critical revision to improve this work.



\section*{Data availability}

The data underlying this article are available in the article and in its online supplementary material.



\bibliographystyle{mnras}
\bibliography{reference_final} 



%
%
%
%

\bsp	
\label{lastpage}
\end{document}